%% file: cbi_instr.tex
\documentclass[useAMS,usenatbib]{mn2e}

\usepackage{graphicx} 
\usepackage{verbatim} 
\usepackage{subfigure}
\usepackage{amssymb,amsmath}
\graphicspath{{./plots/}}

\usepackage{amsmath,amssymb}
\usepackage{natbib}
\usepackage{fancyhdr}
\usepackage{rotating}
\usepackage{setspace}
\usepackage{color}
\usepackage{verbatim}
\usepackage{appendix}
\usepackage{booktabs}
\usepackage[T1]{fontenc}
\usepackage{mathptmx}
\usepackage{rotating}
\usepackage{longtable}
\usepackage{pdflscape}
\usepackage{supertabular}
\usepackage{multirow}
\input{rgb}

\title[The Cosmic Background Imager 2]{The Cosmic Background Imager 2} \author[Angela~C. Taylor et al. ] {Angela~C. Taylor$^{1}$\thanks{E-mail: act@astro.ox.ac.uk}, Michael~E. Jones$^{1}$, James R. Allison$^{1,2}$,  Emmanouil Angelakis$^{3}$, \and J. Richard Bond$^{4}$, Leonardo Bronfman$^{5}$, Ricardo Bustos$^{5,6}$, Richard J. Davis$^{7}$, \and Clive Dickinson$^{7}$, Jamie Leech$^{1}$, Brian S. Mason$^{8}$, Steven T. Myers$^{9}$, Timothy J. Pearson$^{10}$, \and Anthony C. S. Readhead$^{10}$, Rodrigo Reeves$^{10,11}$,  Martin C. Shepherd$^{10}$ \and  and Jonathan L. Sievers$^{4}$\\
\\$^{1}$University of Oxford, Department of Physics, Keble  Road, Oxford, OX1 3RH, UK
\\$^{2}$Sydney Institute for Astronomy, School of Physics A28, University of Sydney, NSW 2006, Australia
\\$^{3}$Max-Planck-Institut fur Radioastronomie, Auf dem Hugel 69, 53121 Bonn, Germany
\\$^{4}$Canadian Institute for Theoretical Astrophysics, University of Toronto, ON M5S 3H8, Canada
\\$^{5}$Departamento de Astronom\'ia, Universidad de Chile,  Casilla 36-D, Santiago, Chile
\\$^{6}$Departamento de Astronom\'ia, Universidad de Concepci\'on, Casilla 160-C, Concepci\'on, Chile
\\$^{7}$Jodrell Bank Centre for Astrophysics, School of Physics \& Astronomy, The University of Manchester, Oxford Road, Manchester, M13 9PL, UK
\\$^{8}$National Radio Astronomy Observatory, 520 Edgemont Road, Charlottesville,  VA 22903, USA
\\$^{9}$National Radio Astronomy Observatory, Socorro, NM 87801, USA
\\$^{10}$Cahill Center for Astronomy and Astrophysics, Mail Code 249-17, California Institute of Technology, Pasadena, CA 91125, USA
\\$^{11}$Departamento de Ingenier\'ia El\'ectrica, Universidad de Concepci\'on, Concepci\'on, Chile}
\begin{document}

\date{Accepted 2011 August 18.  Received 2011 August 18; in original form 2011 July 7}

\pagerange{\pageref{firstpage}--\pageref{lastpage}} \pubyear{2011}

\maketitle

\label{firstpage}

\begin{abstract}
  We describe an upgrade to the Cosmic Background Imager instrument to
  increase its surface brightness sensitivity at small angular
  scales. The upgrade consisted of replacing the thirteen 0.9-m
  antennas with 1.4-m antennas incorporating a novel combination
  of design features, which provided excellent sidelobe and spillover
  performance for low manufacturing cost. Off-the-shelf spun primaries
  were used, and the secondary mirrors were oversized and shaped
  relative to a standard Cassegrain in order to provide an optimum
  compromise between aperture efficiency and low spillover
  lobes.  Low-order distortions in the primary mirrors were compensated
  for by custom machining of the secondary mirrors. The
  secondaries were supported on a transparent dielectric foam cone to
  minimize scattering. The antennas were tested in the complete
  instrument, and the beam shape and spillover noise contributions
  were as expected. We demonstrate the performance of the telescope
  and the inter-calibration with the previous system using
  observations of the Sunyaev-Zel'dovich effect in the cluster Abell
  1689. The enhanced instrument has been used to study the cosmic
  microwave background, the Sunyaev-Zel'dovich effect and diffuse
  Galactic emission.

\end{abstract}


\section{Introduction}\label{Introduction}

The Cosmic Background Imager (CBI) \citep{Padin:2002} was a 13-element
co-mounted interferometer operating at 26-36~GHz, designed primarily to
observe the power spectrum of fluctuations in the cosmic microwave
background (CMB) on angular scales of 5 arcmin to 1 deg (multipoles
$\ell \sim 400$ to $\ell \sim 3500$). Between 2000 January and 2005
April, the CBI operated from the Chajnantor Plateau, Chile at an
altitude of 5100~m and during this period it made observations
of the CMB power spectrum in both intensity and polarization
\citep{Padin:2001,Mason:2003, Pearson:2003, Readhead:2004a,Readhead:2004b,
  Sievers:2003,Sievers:2007, Sievers:2009}. In addition it was also used to make
observations of the Sunyaev-Zel'dovich (SZ) effect in a sample of
low-redshift ($z\le0.1$) clusters \citep{Udomprasert:2004}, and
measurements of `anomalous' microwave emission from dust in a range of
Galactic objects
\citep{Casassus:2004,Casassus:2006,Casassus:2008,Hales:2004,Dickinson:2006,Dickinson:2007}.

\begin{figure*}
\centering {\includegraphics[width=0.45\textwidth]{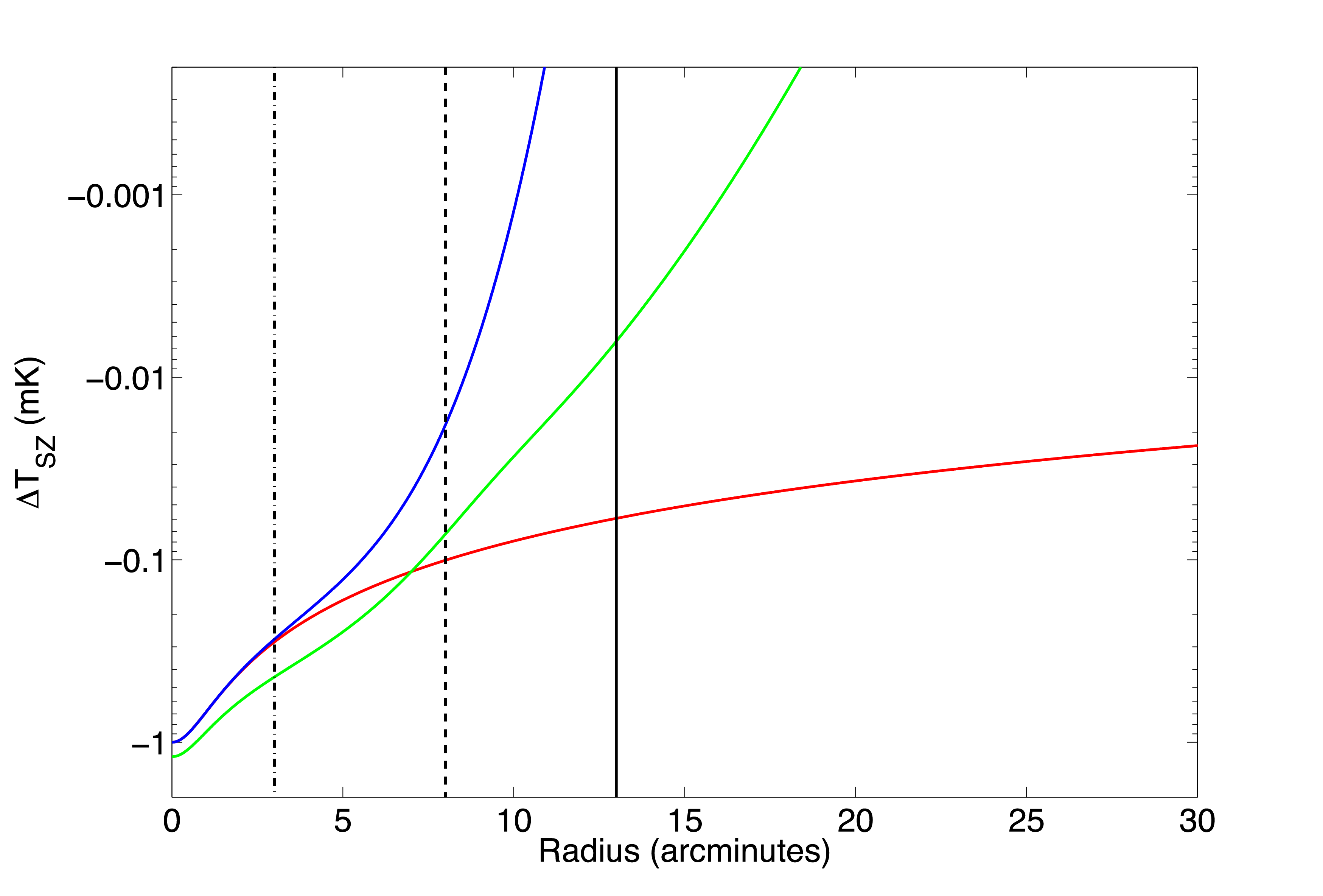}}
\centering {\includegraphics[width=0.45\textwidth]{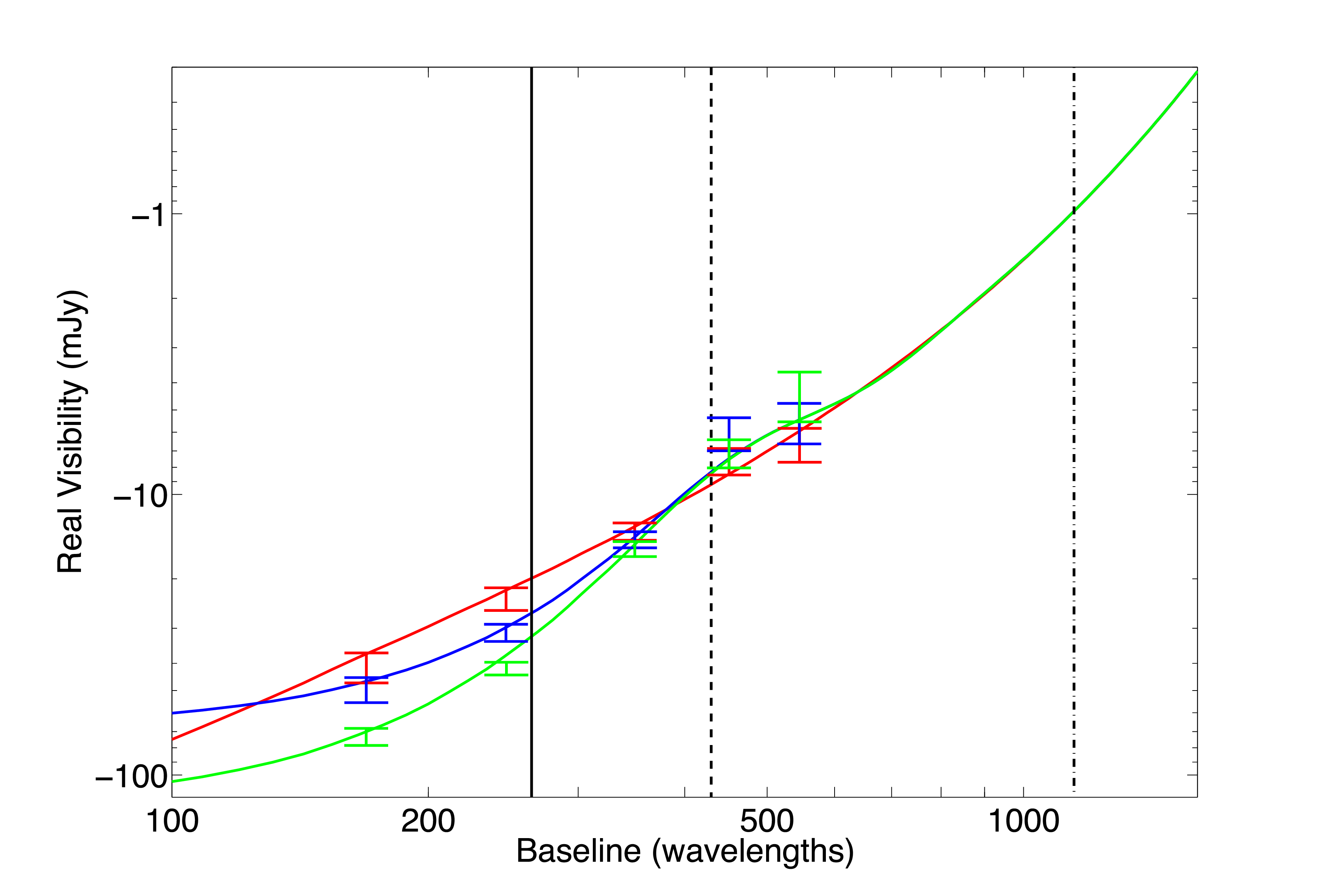}}
\caption{\emph{Left:} The change in thermodynamic temperature relative
  to the CMB due to the thermal SZ effect from three clusters, at $z =
  0.15$, with different large-scale properties. The vertical lines
  represent radii at $r_{2500}$ (\emph{dashed-dotted}),
  $r_{500}$ (\emph{dashed}) and $r_{200}$ (\emph{solid}). The profiles
  are (red: isothermal beta model) $\Delta T_1 = \Delta T_0 (1 +
  (r/r_1)^2)^{1/2 - 3\beta/2}$ with $r_1 = 0.75 \,
  \rm{arcmin}$, $\beta = 0.7$ and $\Delta T_0 = 1.0 \,\rm{mK}$; (blue:
  suppressed large scales) $\Delta T_2 = \Delta T_1
  \exp(-(r/r_2)^4)$ with $r_2 = 7 \, \rm{arcmin}$; and
  (green: additional large scale component) $\Delta T_3 = \Delta T_2 +
  0.2 \exp(-(r/r_2)^2) \, \rm mK $. \emph{Right:} The
  corresponding visibility amplitude with 1.4-m diameter antennas as
  a function of baseline length $u$, at $\nu = 31\,$GHz. The vertical
  lines represent baseline lengths corresponding to the three radii
  (e.g., $u_{2500} = 1/r_{2500}$). The error bars represent
  typical data from the CBI2 experiment, including both thermal noise
  and intrinsic CMB components.}
\label{figure:large_scale_sensitivity}
\end{figure*}

These observations were made using antennas 90~cm in diameter.  In
2005 -- 2006, the CBI was upgraded to larger 1.4-m antennas (`CBI2') to
increase the effective collecting area and to allow observations at
higher resolution without compromising surface brightness sensitivity.
Observations with the CBI2 continued until 2008 June, after which its site
and mount were used for the QUIET experiment
\citep{Quiet:2010}.  During this period the CBI2 completed a programme of
observations of diffuse Galactic emission, the CMB power spectrum and
targeted SZ clusters \citep [and
  further papers in preparation]{Dickinson:2009, Dickinson:2010,
  Castellanos:2010, Vidal:2011}. In this paper we describe the antenna
design that was used in the CBI2 upgrade. We summarize the main
science goals for the upgrade and present commissioning results that
confirm its effectiveness. We also present a combined analysis of an
SZ detection in the cluster A1689. This cluster was observed both with
the original CBI (hereafter `CBI1') and with the upgraded CBI2 and
allows us to demonstrate both the inter-calibration of the two
instruments and the benefit of measuring the SZ decrement with the larger CBI2 antennas.

\section{Science motivation}

The angular scales to which an interferometer is sensitive are set by
the lengths of the baselines between the antennas, with longer
baselines responding to finer-scale information in the sky
brightness. However, for a fixed antenna size, the sensitivity of a
baseline to extended sources decreases rapidly as the baseline is
lengthened. In the Rayleigh-Jeans limit, the temperature sensitivity $\Delta T$ is given approximately by
\begin{equation}
\Delta T = \lambda^{2} \Delta S / (2kf\Omega) ,
\end{equation}
 where $\Delta S$
is the flux density (point source) sensitivity, $\Omega$ is the solid angle of the main lobe
of the synthesized beam, and the filling factor, $f$, is the fraction
of the synthesized aperture that is filled with antennas. This is
simply a modification of the Rayleigh-Jeans equation to reflect the
fraction of photons captured instantaneously by the aperture -- the
exact temperature sensitivity as a function of angular scale will
depend of the configuration of the antennas within the synthesized
aperture. Increasing the resolution of an interferometer without
losing brightness sensitivity thus requires that either the number of
antennas be increased, or the antenna size be increased, in order to
maintain the filled fraction of the synthesized aperture. If the
number of baselines is fixed, and the antennas are not changed,
lengthening the baselines results in an increase in integration time
to reach the same temperature sensitivity proportional to the fourth
power of the baseline length.

The primary goal of the CBI2 upgrade was to increase the temperature
sensitivity of the instrument on its longer baselines of 3 -- 5.5~m,
i.e., corresponding to angular scales of 6 -- 12 arcmin, on which the
CBI1 array was not well filled. Improved sensitivity on these longer
baselines would provide significantly improved observations of the SZ
effect in massive galaxy clusters. In CBI1 SZ observations, the
shortest baselines were heavily contaminated by primary CMB
anisotropies, while the longer baselines lacked thermal
sensitivity. Moderately massive clusters typically have virial radii
of $\sim$\,2\,Mpc, which at a redshift of $z \sim 0.15$ corresponds to
an angular size of $\sim 12$\,arcmin. This is well-matched to the new
CBI2 array, which is thus able to measure the cluster gas out to the
outskirts of the clusters with significantly less contamination from
primary CMB fluctuations than was the case for CBI1.

The motivation to concentrate on measuring the SZ effect out to the
virial radius in complete samples of clusters was driven by the need to
further understand the X-ray--SZ and weak-lensing--SZ scaling
relations in support of SZ survey experiments. The SZ effect measures
the Comptonization parameter, $y = \int kT_{\rm e}/(m_{\rm e}c^2) {\rm
  d}l$, which is proportional to the electron pressure integrated
along the line of sight. SZ surveys are designed to measure the
integrated SZ effect, $Y = \int {y\,\rm{d}\Omega}$, providing
empirical measurements of the cluster co-moving SZ luminosity function
${\rm d}N/{\rm d}Y$. However, in order to relate these measurements to
cosmology via the cluster mass function, ${\rm d}N/{\rm d}M$, a
well-calibrated relationship between $Y$ and the total mass $M$ is
required. This can be achieved by combining SZ measurements of known
clusters with X-ray and weak lensing data, along with modelling that
accurately describes the distribution of the cluster components (dark
matter, gas and galaxies) in a way that can be constrained by the
observational data. There have been a number of recent measurements of
the scaling between the integrated SZ effect and the total mass, from
both hydrostatic \citep{Benson:2004,Bonamente:2008} and gravitational lensing
\citep{Marrone:2009} estimates. However these relationships have
generally only been obtained out to relatively small radii ($\sim
200-400$\,kpc) and observations with experiments such as CBI2, APEX-SZ
\citep{Schwan:2003} and AMIBA \citep{Ho:2009} are expected to provide
constraints out to a few Mpc ($\sim r_{200}$, where $r_{x}$ denotes
the radius within which the average density is $x$ times the critical
density).

Figure\,\ref{figure:large_scale_sensitivity} illustrates why
measurements at large angular scales relative to the core of the
cluster are important in determining true SZ profiles. It shows the
thermal SZ effect for toy models of three clusters at $z = 0.15$, with
similar cores but different large-scale properties. The red curve is a
standard isothermal beta model \citep{Cavaliere:1976,Cavaliere:1978}, 
\begin{equation}
\Delta T = \Delta T_0 (1 +(r/r_{\rm core})^2)^{1/2 - 3\beta/2} 
\label{eqn:isobeta}
\end{equation}
where $\Delta T_0$ is the central temperature decrement, $r_{\rm
  core}$ is the cluster angular core radius and $\beta$ controls the
shape of the profile.  The blue profile has the same radial behaviour
within the cluster centre but has an exponential decline which becomes
significant beyond $r \gtrsim r_{2500}$ (corresponding, for example,
to a declining temperature profile). The green profile has a similar
radial behaviour to the blue, but with an additional additive
large-scale component that produces a larger overall value for the
central SZ signal. The corresponding visibility profiles (as a
function of baseline length in wavelengths $u$) are obtained by taking
the amplitude of the fourier transform of the product of the SZ model,
$\Delta T(r)$, with the primary beam of the interferometer $B(r)$
(here assumed to be a Gaussian of half-power width 30 arcmin). The
signal is converted from temperature to flux density units using the
Planck equation. For a circularly symmetric model this is easiest to
implement using a Hankel transform,
\begin{equation} 
\Delta S(u) = \frac{2k}{c^2}\frac{x^2e^x}{(e^x-1)^2} 2\pi \int_0^\infty \Delta T(r) \, B(r) J_0(2 \pi r u) r {\rm d}r,
\end{equation} 
where $x = h\nu/kT_{\rm CMB}$ and $J_0$ is the zeroth-order Bessel function.

These visibility profiles show that interferometric experiments cannot
distinguish between the SZ profiles on baselines greater than $~400
\, \lambda$. (The same would be true for total power measurements where
the data are spatially filtered on scales greater than the equivalent
angular scale, here about 10 arcmin.) However the integrated SZ flux
density, which corresponds to the total thermal energy in the cluster
and is the quantity which is measured as a proxy for mass in cluster
surveys, varies by almost a factor of two between these cases. It is
therefore important to observe on baselines short enough to
distinguish between different large-scale cluster properties. In the
case of observing frequencies $\sim 30\,$GHz, this requires an
interferometer with baselines smaller than around $4\,$m.

Figure\,\ref{figure:large_scale_sensitivity} also displays the
expected data that would be obtained from CBI2 observations of these
three SZ profiles. The error bars include both a thermal noise
component and a component due to the intrinsic CMB anisotropy, the
latter being significant on baselines shorter than $250\lambda$. The
uncertainty due to the thermal noise integrates down with the square root of
the observing time, whereas the contamination due to the primordial
CMB fluctuations does not. Without additional frequency information
which can distinguish between the CMB and the characteristic spectrum
of the SZ effect, sensitivity to the large-scale SZ effect is
ultimately limited by the primordial CMB fluctuations. The CBI2
baselines however provide an excellent compromise between primary CMB
contamination on the one hand and resolving out of the largest scale
emission on the other.

\section{Upgrade of the CBI antennas}

\subsection{CBI1 antenna design}

The original antenna design for the CBI1 was an on-axis Cassegrain with
a $0.9$-m diameter, $f = 0.33$ primary, and a 155~mm-diameter
hyperboloidal secondary with eccentricity of 1.41
\citep{Padin:2002}. The secondary was supported on a transparent
polystyrene quadrupod, and the whole antenna enclosed in a can rising
to 400~mm above the rim of the primary. This can was designed to
reduce the coupling from the secondary of one antenna to the feed of
the adjacent antenna, and was measured to reduce such coupling from ~
$-90$ dB to $\sim \, -120$~dB. The secondary mirror was oversized, in
the sense that it extended beyond the radius required to reflect a ray
from the feed to the edge of the primary. This is a common feature of
Cassegrain designs, and is intended to increase the aperture
efficiency. By increasing the size of the secondary, the diffraction
beam of the secondary is reduced, resulting in a sidelobe which would
otherwise have missed the primary edge hitting the primary, and
therefore contributing to the main aperture illumination. It also has
the effect however, of providing additional direct ray paths from the
secondary over the edge of the primary, in the direction of the
adjacent antennas. To alleviate this problem, the original antennas
were provided with the shield can, which greatly reduced the spillover
in the backward direction, and redirected the spillover power to the
sky in the general direction of the main beam.

\subsection{CBI2 antenna design} 

The new antenna design for CBI2 was intended to make maximum use of
the physical area of the platform on which the CBI1 antennas were
co-mounted. The largest antenna size that could be accommodated on
the table while still using all 13 antennas was 1.4~m diameter. The
design had to be cost-effective and reasonably quick to implement, and
so was based on a commercially available reflector with a nominal
diameter of 1.37~m and actual maximum diameter 1.41~m (the difference
being due to the roll-off of the surface at the rim). The focal length
was 457~mm, giving an $f$-ratio of 0.33, very similar to the original
CBI1 design. The reflector was fabricated by spinning, in which a
circular aluminium sheet is pressed over a spinning mould with a
roller. The surface is then rolled over a circular tube at the rim,
and finally a circular tube is riveted to the back of the dish at
300~mm radius to provide a mounting point. This method is very quick
and cheap, and measurements with a co-ordinate measuring machine showed that the surface accuracy $\delta x$ of a sample dish was
better than 0.2 mm rms on small scales over most of the dish
surface. However, the dishes also typically had large-scale distortion
of the form $\delta x \propto r^2 \cos(2\theta)$, consistent with the rim
tube being elliptical with a deviation from circular of several
millimetres, forcing the surface out of its paraboloidal shape. This
error was dealt with by modifying the secondary optics as described
below.

\begin{figure}
{\includegraphics[width=0.47\textwidth]{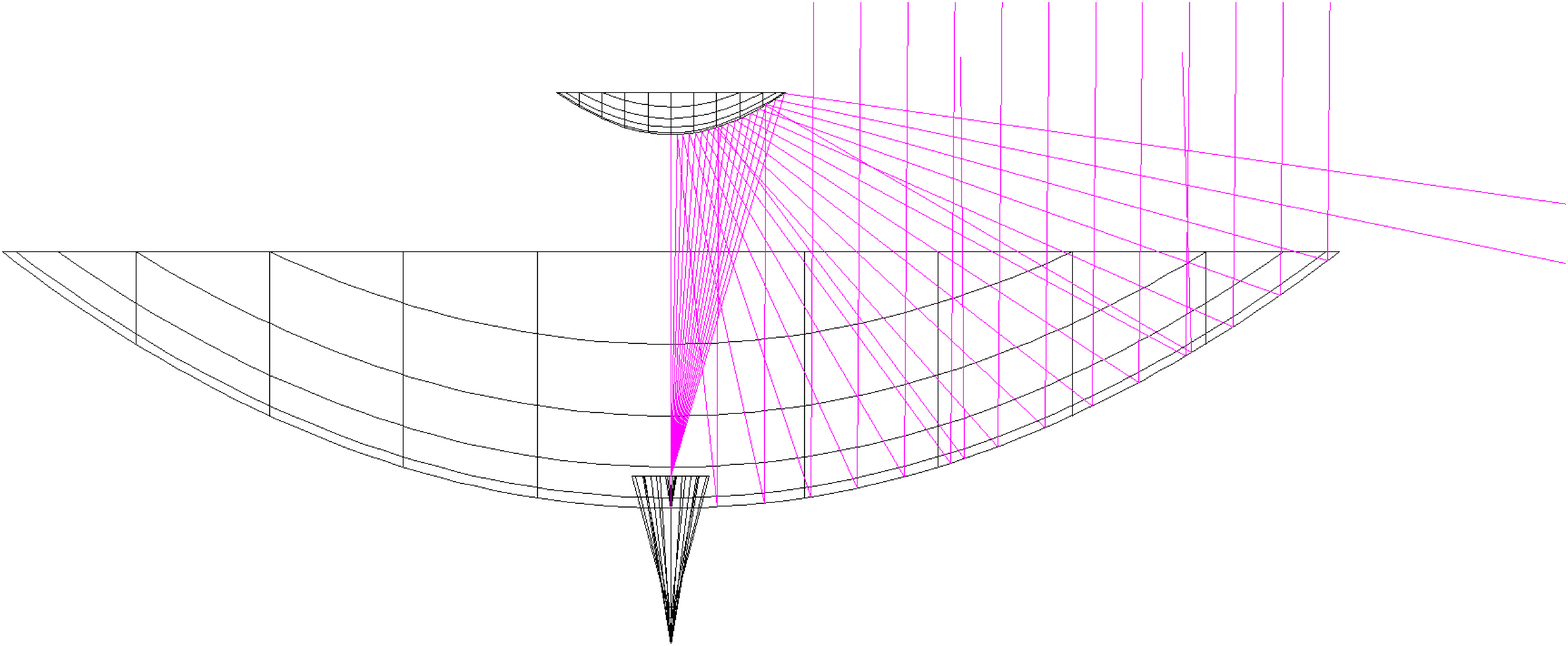}}
{\includegraphics[width=0.4\textwidth]{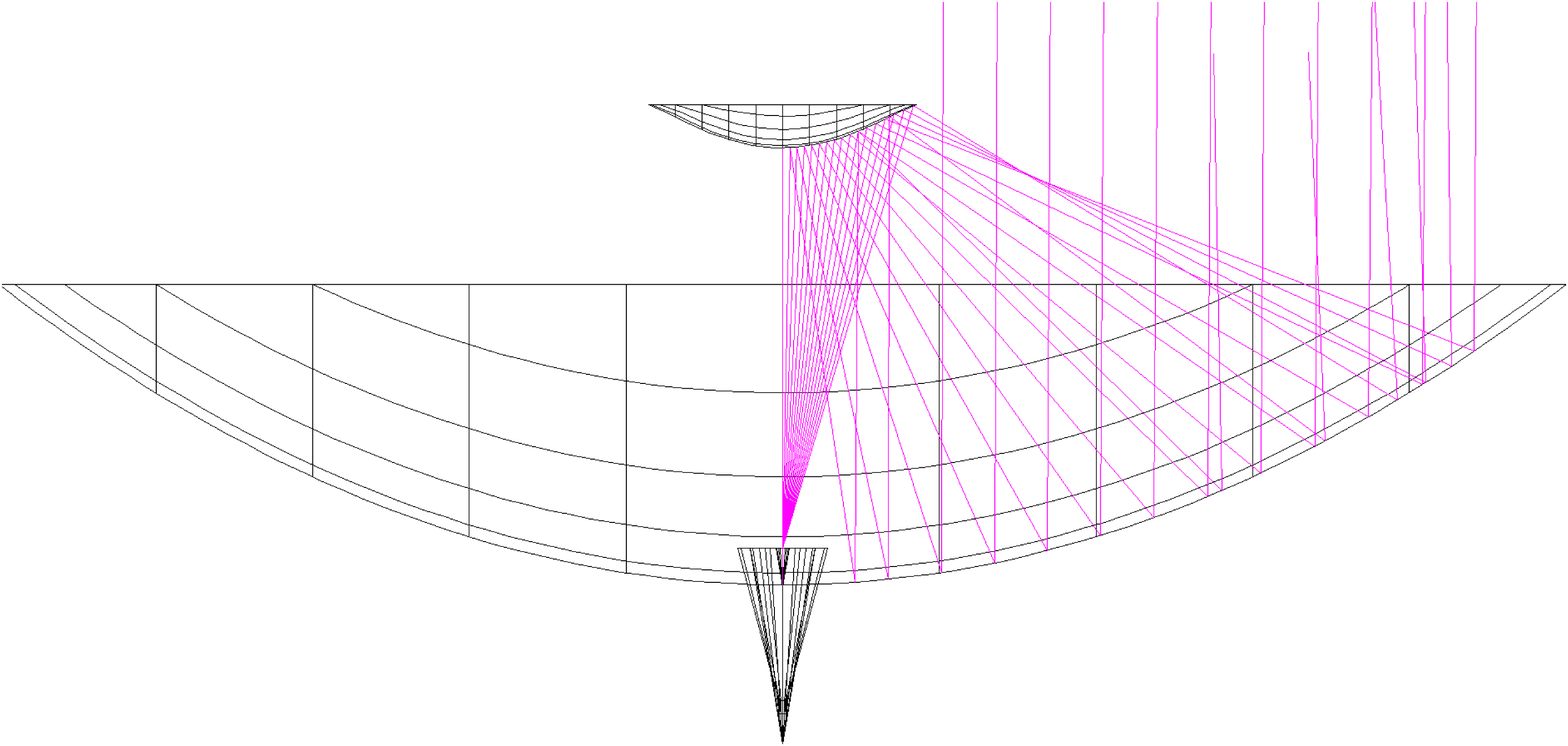}}
\caption{Ray diagrams of the CBI2 antenna with an oversized secondary with (\emph{top}) no reshaping, showing the potential for spillover past the primary and (\emph{bottom}) with the secondary reshaped at large radii to redirect the spillover rays back on to the primary. }
\label{figure:secondary_optics}
\end{figure}

\subsection{Optical design of the CBI2 secondary optics}

\begin{figure}
\centering {\includegraphics[width=0.45\textwidth]{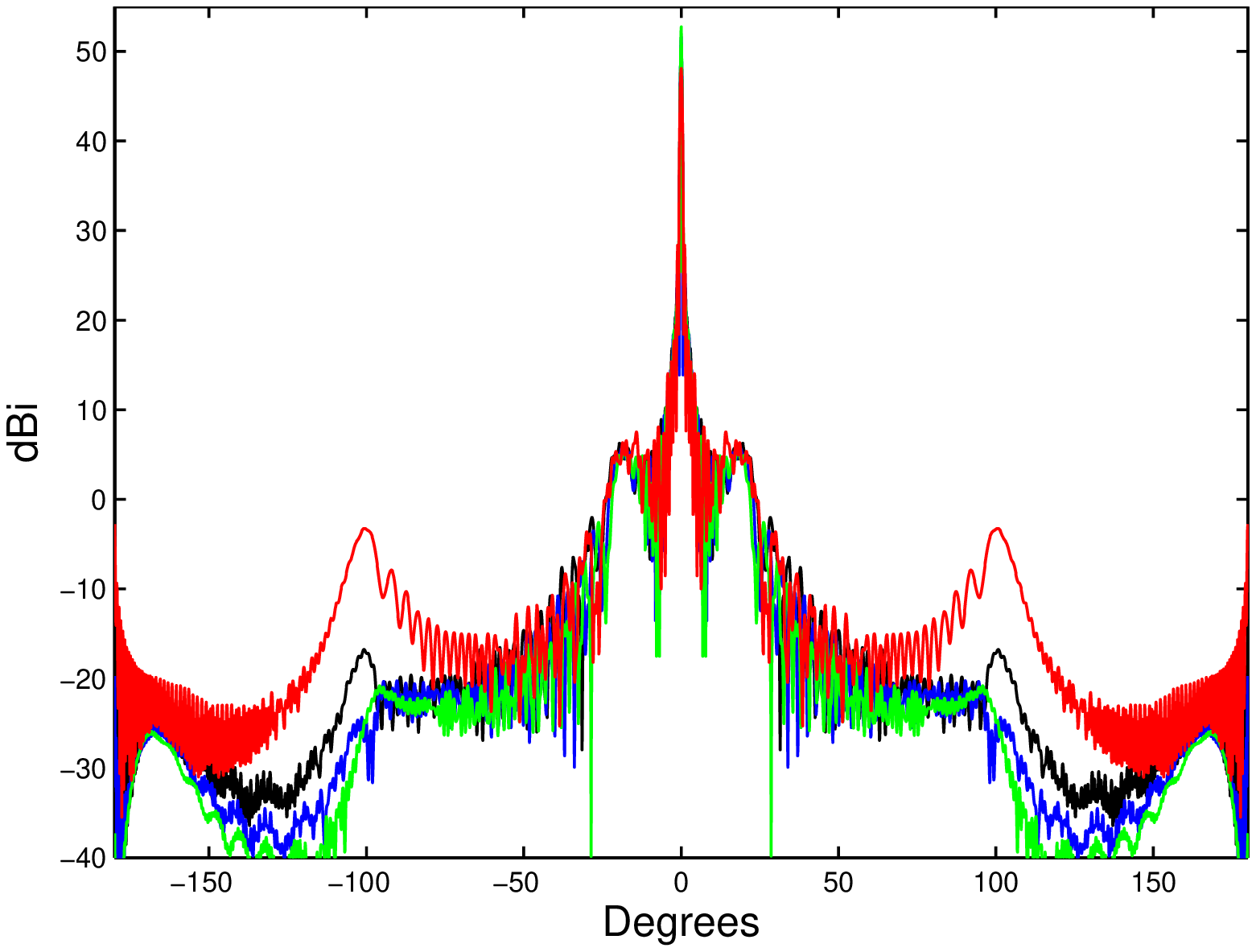}}
\hspace{3cm}
\centering {\includegraphics[width=0.45\textwidth]{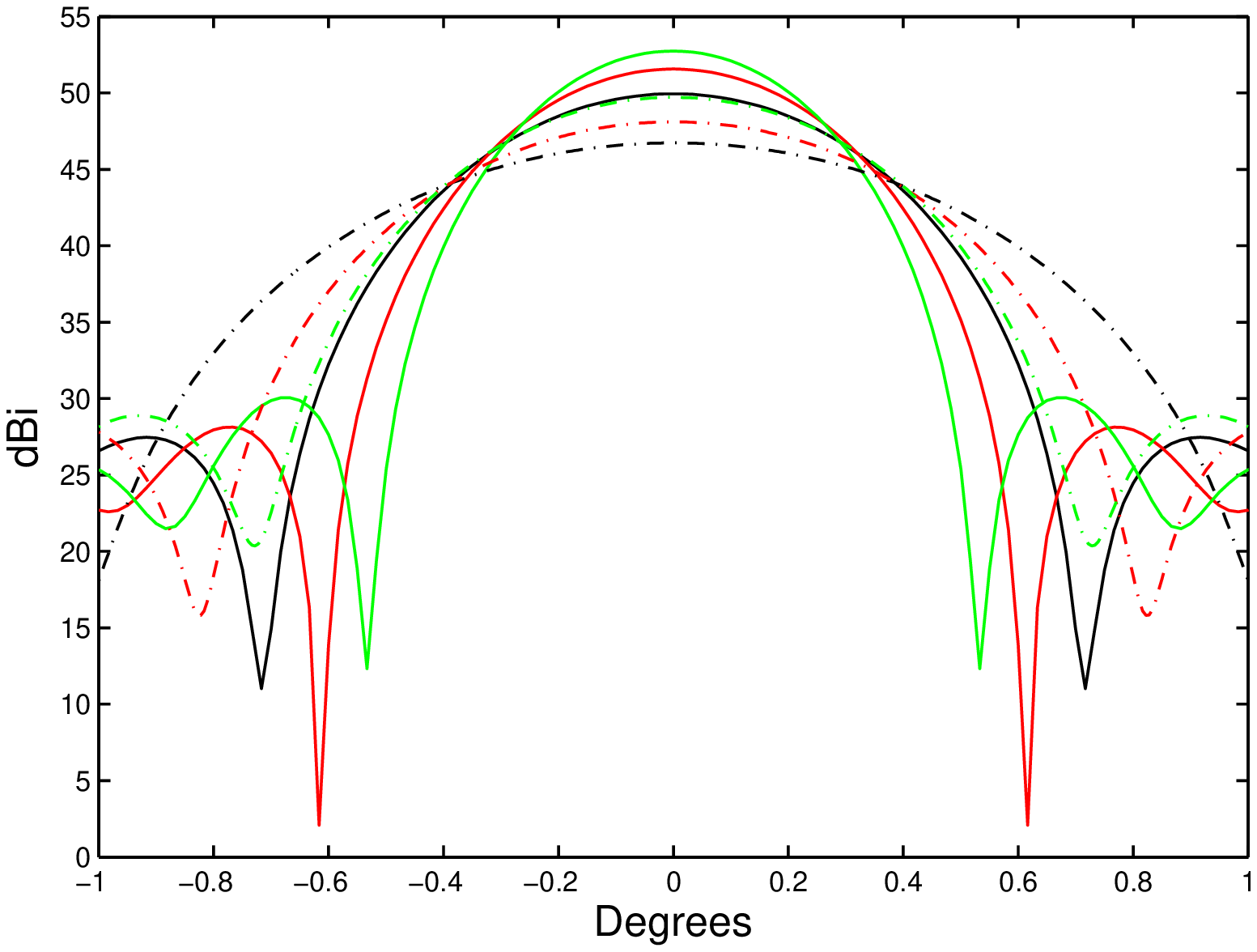}}
\caption{Comparison of calculated beam patterns of the CBI1 and CBI2
  antennas across the observing band of 26-36 GHz. \emph{Top:} Far-out
  beam patterns: Black, 1.4-m antenna at 26~GHz; blue, 1.4-m at
  31~GHz; green 1.4-m at 36~GHz; red 0.9-m antenna at 31~GHz (no
  shield can). Note that the main spillover lobe at ~100 deg is lower
  in all cases for the larger antenna and is negligible at the top of
  the observing band. \emph{Bottom:} Main lobe of the calculated beam
  patterns. Solid lines indicate the 1.4-m antenna. Dashed lines indicate the 0.9-m antenna. In both cases the colour scheme is green, 36~GHz; red, 31~GHz and black, 26~GHz. }
\label{figure:beam_cuts}
\end{figure}

The main competing design drivers were aperture efficiency versus
sidelobe spillover. The existing CBI1 feed horn was modelled using the
\textsc{Corrug}\footnote{SMT Consultancies:
  http://www.smtconsultancies.co.uk/products/\\corrug/corrug.php}
software package and the resulting feed pattern used to illuminate a
model of the primary dish in the \textsc{GRASP9}\footnote{TICRA:
  http://www.ticra.com/what-we-do/software-descriptions/grasp/}
software package. \textsc{GRASP9} enables full physical optics plus
physical theory of diffraction simulations to be done on the complete
optical system.  This method takes into account the fields from both
the surface and edge currents on the reflectors, as well as blockage
and multiple reflections (e.g., in the region of the primary shadowed
by the secondary). In order to minimize spillover without sacrificing
too much aperture efficiency, the CBI2 optics design incorporated a
secondary mirror which was both oversized and reshaped
\citep{Holler:2008}. In order to find the optimum balance between
aperture efficiency and sidelobe spillover, the secondary mirror size
was increased from the nominal ray optics size, increasing the
aperture efficiency due to diffraction effects, until the increasing
blockage began to reduce the efficiency again. The edge of the
secondary was then reshaped by adding a quadratic term to the
hyperboloid, starting at a point near the ray optics illumination edge
(i.e., the point where an on-axis ray striking the edge of the primary
would strike the secondary). This has the effect of directing
radiation closer inwards on the primary than would otherwise be the
case. The ray traces are shown in Figure \ref{figure:secondary_optics}
(ray traces are not accurate modelling tools for cases such as this
where the antenna properties are dominated by diffraction effects, but
are useful to visualize the design concepts). Both the starting point
and amplitude of the quadratic term were adjusted to achieve a
reasonable compromise between aperture efficiency and sidelobe
level. In addition, the distance between the primary and secondary
mirrors was adjusted from the geometric optics value in order to
maximize the forward gain, to take account of the fact that the optics
is all in the near field of the feedhorn. This resulted in a shift of
the secondary position by 5 mm towards the primary and an improvement
of the forward gain by about 25 per cent.

\begin{figure*}
\centering 
\includegraphics[trim= 20mm 15mm 0mm 0mm,clip,width=0.45\textwidth]{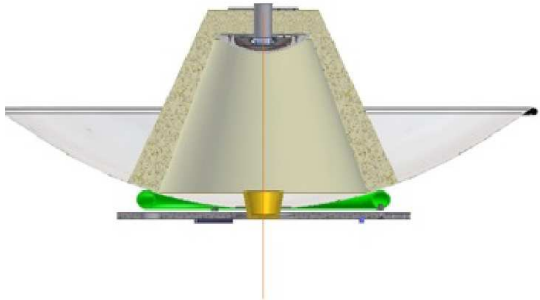}
\includegraphics[width=0.4\textwidth]{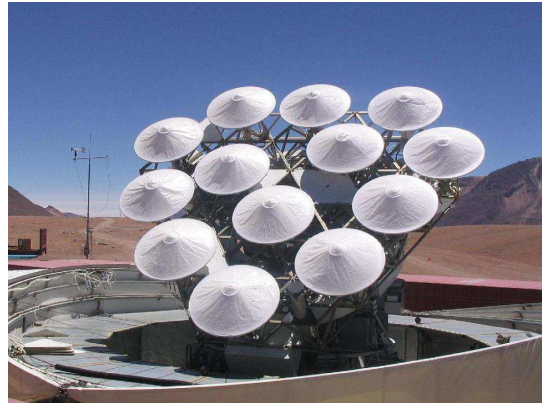}
\caption{{\em Left:} A cross-section drawing of the CBI2 dish and foam
  cone assembly. {\em Right:} The thirteen new CBI2 antennas mounted
  on the triaxial mount. The antennas are protected from the weather
  by individual woven polyethylene sheet radomes.}
\label{figure:array_config}
\end{figure*}

The central 14~mm of the secondary was reshaped into a cone of
semi-angle 86 deg, such that rays from the feed close to the axis are
not reflected directly back on to the cryostat window, causing standing
waves. Figure \ref{figure:beam_cuts} shows cuts in one plane through
the calculated beam patterns at 26, 31 and 36~GHz. Also shown for
comparison is the calculated pattern for the 0.9-m antennas (without
shield can) at 31~GHz. It can be seen that the spillover lobes are
reduced by around 20~dB compared to the previous design, and that
there is essentially no spillover lobe at the top end of the observing
band.

The main beam plots in Figure \ref{figure:beam_cuts} also show that
the forward gain of the CBI2 antenna is 3 dB greater at each frequency
compared to the old CBI1 design, i.e., the effective aperture is
bigger by a factor of two. This is smaller than the nominal area
increase (($1.4/0.9)^2 = 2.4$) due to the under-illumination (or
steeper edge taper) caused by the reshaping, which was necessary in order to keep the spillover lobes to a minimum.
 
\subsection{Correction for primary asymmetry}

The optical design was developed using an ideal model of the 1.4~m
primary dish. However, as described earlier, the manufacturing process
used to make the primary dishes introduces a large-scale deformation,
as a result of using a non-perfectly circular reinforcing ring at the
rim of the dish. This results in a non-circular beam-pattern with
reduced forward gain.  To compensate for this effect the surface
profile of each CBI2 primary dish was measured along several circular
tracks at differing radii. For all but one of the dishes it was
possible to get a good fit to the distortion using just the
second-order Zernicke polynomial $Z_2^2 = r^2 \cos(2\theta)$ (the
other dish also required an inclusion of a $Z^2_3 = r^2 \cos(3\theta)$
term). By adding the appropriate Zernicke polynomial to each of the
secondary antenna profiles it is possible to cancel out the effect of
the distortion \citep{O'Sullivan:2007}. This was done for each dish in
turn, using the \textsc{Zemax}\footnote{Zemax Development Corporation,
  http://www.zemax.com/} optics modelling package in order to
determine the amplitude of the polynomial correction needed to
maximize the Strehl ratio in a ray optics simulation.  \textsc{GRASP9}
simulations of each of the antennas in turn were made using the
appropriately corrected secondary to verify the design. The tolerance
of the beam shape to focus position was also modelled, as the
corrected optics were noticeably less tolerant to focussing errors
than the ideal optics. The resulting secondary mirror designs
(oversized, reshaped and with an appropriate Zernike polynomial added)
were then machined from solid aluminium using a CNC milling machine,
with care taken to indicate the axis of the polynomial on the
secondary such that it could be aligned with that of the corresponding
primary dish.

\subsection{Antenna assembly}

The new antennas were mated to the existing CBI1 receivers using an
existing mounting plate that in the original design mounted the
receiver, primary mirror and shield can. As in the original design, in order to avoid introducing scattering in the beam, the secondary mirror was supported using a transparent dielectric material.  We used a hollow cone of Plastezote\footnote{Zotefoams plc,
  http://www.zotefoams.com/pages/en/datasheets/ld45.htm} LD45, an expanded
low-density polyethylene foam (the `45' refers to the density in $\rm
kg\,m^{-3}$). Samples of the LD45 were tested in the lab for both their
thermal and electrical properties. Dielectric loss was unmeasurably
small at 30 GHz -- scaling from the volume fraction of the foam the
expected value of the loss tangent would be $\tan \delta = 2 \times
10^{-5}$ -- and the dielectric constant was measured as $\epsilon_r =
1.06$, which is equal to the dielectric constant of solid polyethylene
diluted by the solid volume fraction of the foam. The coefficient of thermal expansion
was however significant at about $5 \times 10^{-5}\, \rm
K^{-1}$. Given a possible temperature range of $>20 \, \rm K$ on site
and the height of the cone of 400 mm, the resulting change in focus
position approaches the maximum tolerance of the design, at around
$\pm 0.5 \, \rm mm$. Care was therefore taken on assembly to ensure
that the nominal focus position was achieved at the mid-range of
expected ambient temperatures (about $-5^{\circ} \, \rm C$).  Each
secondary mirror was attached to a foam lid using a metal plate screwed in
to the back of the mirror, and machined alignment jigs were used to
hold the mirror in place while the lid was glued to the cone. The cone
was assembled from sections bandsawed out of ~100-mm thick LD45 sheet,
glued together on joints perpendicular to the optical axis of the
antenna using the impact adhesive Evostik TX528\footnote{Bostik, http://www.bostik.co.uk/diy/product/evo-stik/TX528/9}. The adhesive joints were very much
thinner than a wavelength, and tests in waveguide showed that they had
negligible attenuation or reflection at 30~GHz. A diagram of the
antenna assembly is shown in Figure \ref{figure:array_config}.
 
\begin{figure*}
\includegraphics[width=0.25\textwidth]{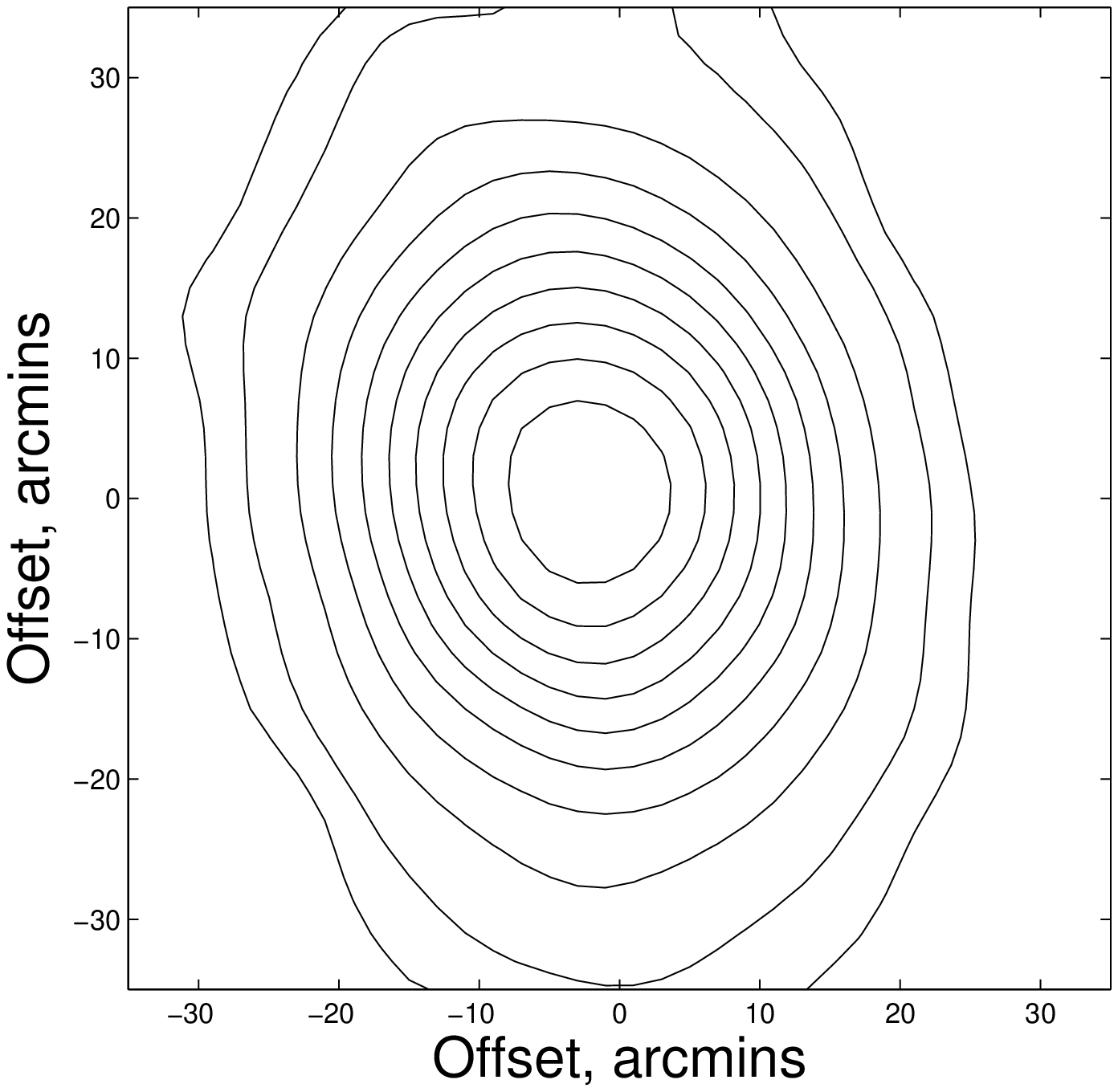}
\includegraphics[width=0.25\textwidth]{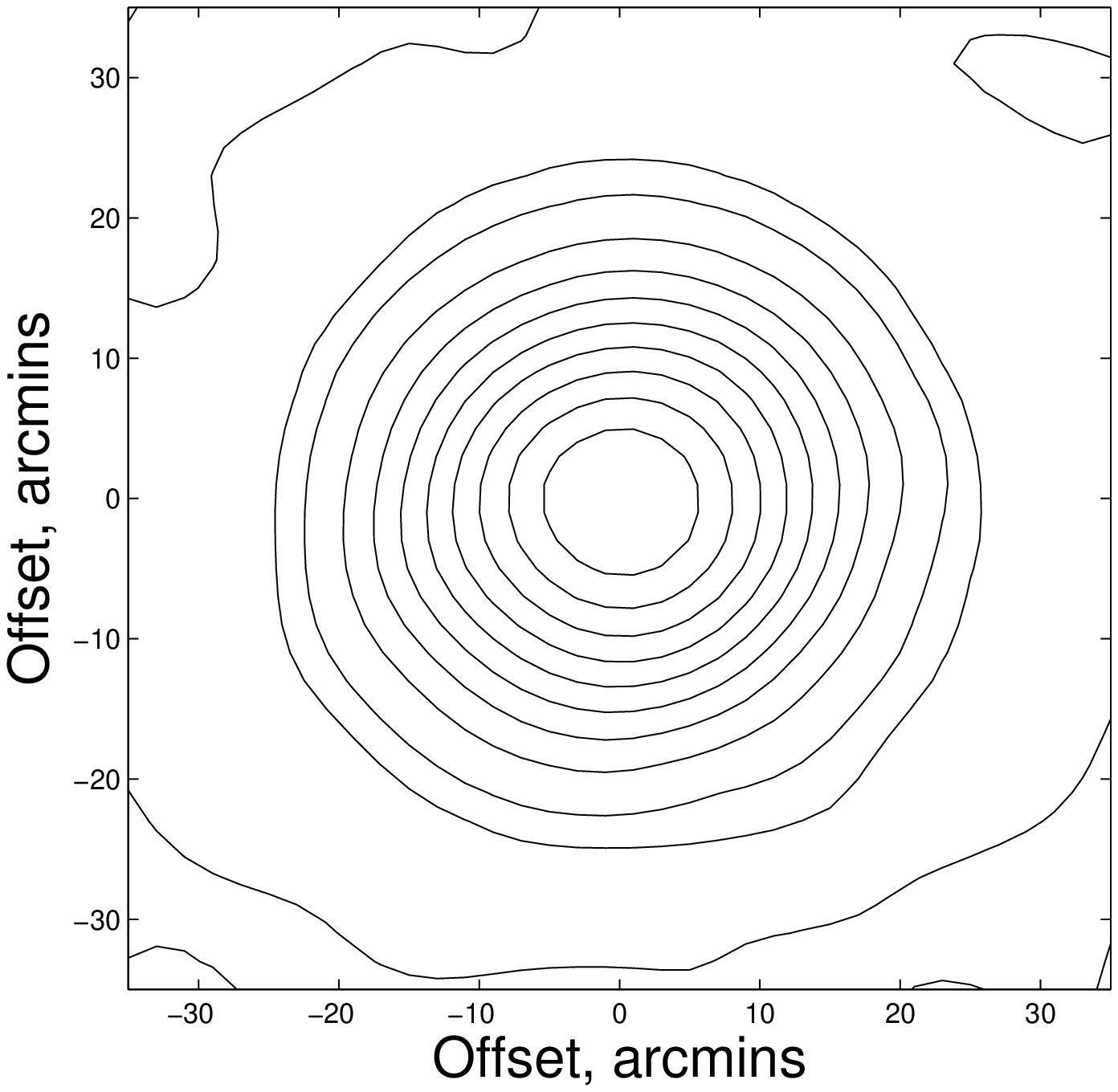}
\includegraphics[angle=-90,trim=0mm 0mm 0mm 10mm,clip,origin=rb,width=0.35\textwidth]{Figure5_right.eps}

\caption{The effect of the correction for primary astigmatism on the
  beam shape of the CBI2 antennas. {\em Left:} Beam pattern measured
  from an observation of Jupiter in a single, 1~GHz frequency channel
  centred at 31.5~GHz for a single antenna with typical primary
  astigmatism but fitted with a symmetric subreflector. {\em Middle:}
  The beam pattern from the same antenna fitted with its individually
  corrected subreflector. Contours are 5, 10, 20 ... 90 per cent of
  the beam peak. Individual antenna patterns are solved for from the
  visibility data for the whole array as it is scanned over the
  source.  {\em Right: }The measured profile of the CBI2 beam at
  31.5~GHz. The scatter points are the measured data from all the CBI2
  baselines in a single, 1~GHz frequency channel centred at
  31.5~GHz. The solid line is the beam simulated in GRASP9, and the
  dashed line is a Gaussian of FWHM 27.8~arcmin (the same FWHM as the
  calculated beam), which fits the beam well within the half-power
  points. }
\label{figure:deformed_dishes}
\end{figure*}

\section{Re-commissioning tests}

All thirteen CBI2 antennas were assembled on-site in Chile prior to
mounting on the CBI platform (Figure \ref{figure:array_config}). The pointing of each antenna was
assessed by making 5-point observations of bright calibrators such as
Jupiter and Tau A. Here the instrument is first pointed on-source, and
then off-source at 4 pointings where the amplitude should be half of
the total signal from the source. Since the CBI2 is a co-mounted
interferometer the pointing errors associated with any individual
antenna have to be separated by modelling the response of each
baseline to the combined pointing error of its pair of antennas and
solving for the individual antenna pointing errors. These differential
pointing errors were corrected for by placing shims under the relevant
mounting feet of each antenna. The residual individual pointing errors
after this process were typically $\sim 0.5$ arcmin.

The primary beam of the new system was measured using observations of
Jupiter on a grid of $11 \times 11$ pointing centres spaced by 7 arcmin (i.e.,
covering offset positions of $\pm$ 35 arcmin in azimuth and
elevation). The integration time per pointing was 45 s. The resulting
beam patterns for one of the CBI2 antennas both with and without a
corrected secondary mirror are shown in
Figure \ref{figure:deformed_dishes} and clearly show the improvement in
the circularity of the beam when using the corrected
secondary. Figure \ref{figure:deformed_dishes} also shows the measured
radial profile of the beam at 31.5~GHz along with the simulated
\textsc{GRASP9} beam. The measured beam can be fitted to the
half-power points with a Gaussian model with a FWHM of 27.8~arcmin at
31.5~GHz, also shown in Figure \ref{figure:deformed_dishes}.

\begin{figure}
\centering
\includegraphics[width=0.45\textwidth]{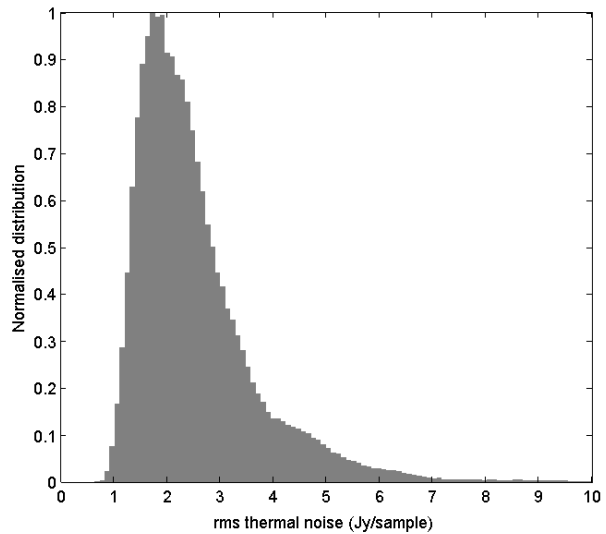}
\caption{The distribution of mean rms noise per 4.2~s sample for all the CBI2
  baselines. The expected value for the nominal system temperature is
  1.9~Jy per sample; the tail to higher values reflects the small fraction of
  antennas with higher than nominal noise.}
  \label{figure:noise_histogram}
\end{figure}

To calculate the expected thermal noise of the new CBI2 array we
assume an effective antenna collecting area of 0.8\,m$^{2}$, effective
bandwidth per channel of 0.85~GHz, correlator accumulation time of
4.2\,s, a nominal system temperature of 30~K and a system efficiency
(due to non-flat passbands, phase errors etc.) of 90 per cent. This
gives an expected rms thermal noise of 1.9\,Jy per sample, or 3.9 Jy s$^{1/2}$. In order to measure the
actual sensitivity a series of blank field observations was made. To
reduce the contribution from ground spill the observations were taken in
differenced mode i.e., with a lead and trail field observed at the same
declination as the main field but separated by 8 minutes in right
ascension.  The resulting sensitivity was then found by calculating
the mean rms of the real and imaginary parts of the visibilities after subtraction
of the lead and trail fields.  Figure\,\ref{figure:noise_histogram}
shows a histogram of the measured noise from CBI2 observations of
blank fields over the period 2007 April 1 to 2008 May 10.  Each datum
was generated from 23 samples of the correlator
output. The measured peak in the rms noise histogram lies at the expected value, but with a tail in the distribution to larger noise
values due to baselines containing receivers with higher than nominal
system temperatures.

\section{SZ detection of A1689 and comparison with CBI1}

As a final check on the effectiveness of the CBI2 upgrade we present
an analysis of an SZ detection in the cluster A1689 \cite[$z =
  0.1832$,][]{Struble:1999}, a hot, massive cluster with a virial mass
$M_{vir} \sim 1-1.5 \times 10^{15} h^{-1}\,\rm{M}_{\odot}$
\citep[]{Limousin:2007, Umetsu:2008, Lemze:2009, Peng:2009} and
an average emission weighted gas temperature of $T_{\rm{ew}} \sim
10.5\,\rm{keV}$ \citep[]{Lemze:2008, Cavagnolo:2009,
  Kawaharada:2010}. This cluster had previously been observed with the
CBI1 and thus serves as a useful check on the cross-calibration of the
two instruments. It also highlights the improved sensitivity of the CBI2
for SZ observations. To allow direct comparison of our observations
with measurements we also fit the combined dataset from the CBI1 and
CBI2 to a single isothermal beta model.

\subsection{Observations and data}

\begin{figure*}
\centering
\includegraphics[trim = 0 0 55 0,clip,width = 0.36\textwidth]{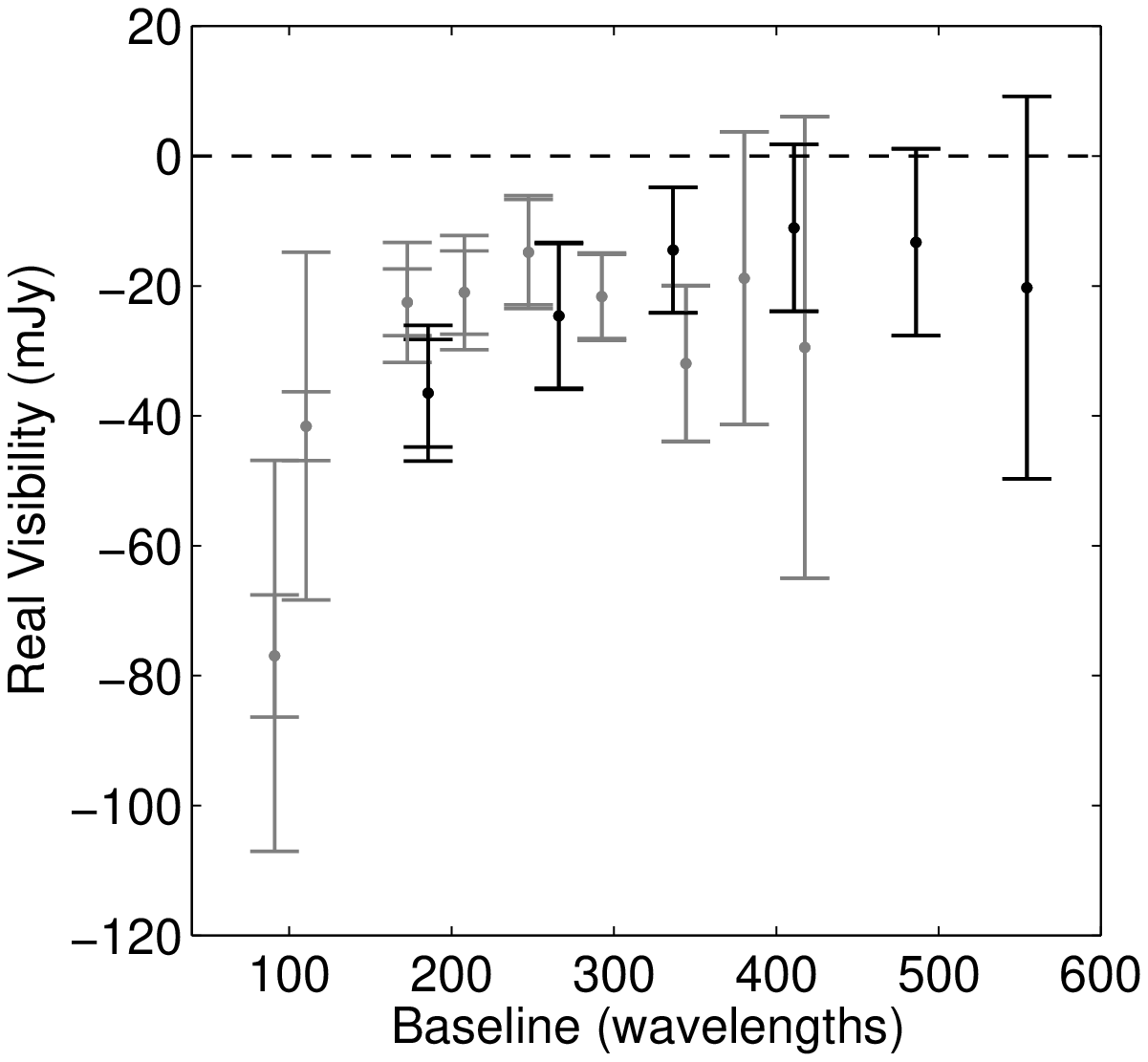}
\includegraphics[origin=c,angle=-90,trim = 0 0 0 0,width = 0.3\textwidth]{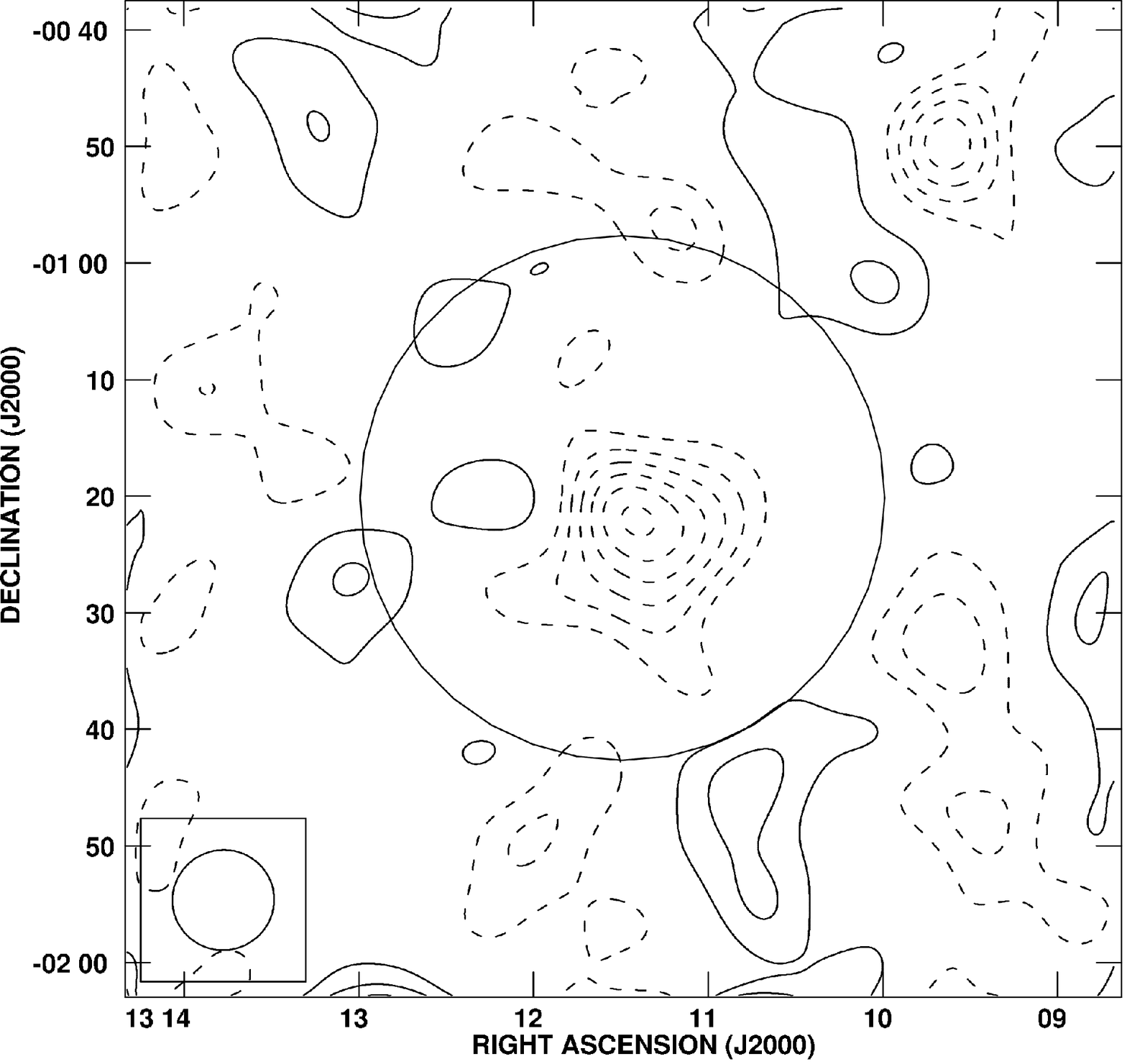}
\includegraphics[origin=c,angle=-90,trim = 0 0 0 0,width = 0.3\textwidth]{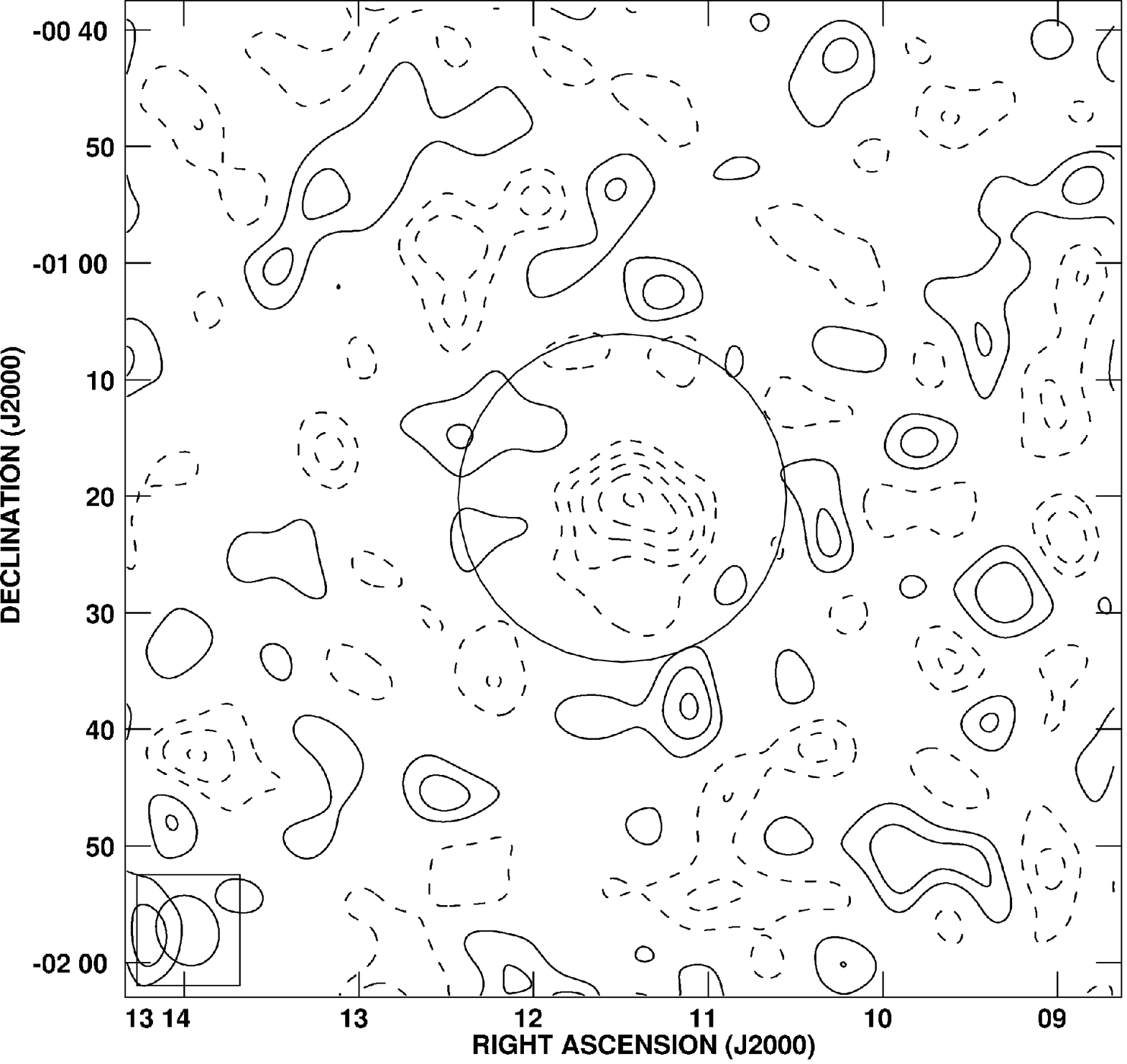}

\caption{\emph{Left:} The real part of the visibility as a function of
  baseline for observations of A1689 using the CBI1 (\emph{grey
    errorbars}) and CBI2 (\emph{black errorbars}). The short errorbars
  represent the $1\,\sigma$ noise from the variance in the data, and
  the long errorbars represent the total statistical uncertainty
  including the intrinsic CMB anisotropy. There is about 50 per cent
  more useable data in the CBI1 observations than in the CBI2
  observations. \emph{Middle and Right:} CLEANed full-resolution maps
  of A1689 using the CBI1 and CBI2 arrays respectively. The rms  noises on each map are 5.9\,mJy\,beam$^{-1}$ for CBI1 and
  6.1\,mJy\,beam$^{-1}$ for CBI2.  The contours are multiples of 6\,mJy\,beam$^{-1}$. The full-width at half-maximum of the synthesized beam is
  shown in the bottom left-hand corner of each map ($8.6\times 8.7$\,arcmin for
  CBI1 and $5.3\times 6.0$\,arcmin for CBI2). The full-width at half-maximum of
  the primary beam, at $\nu = 31\,\rm{GHz}$, is shown as a dark
  circle centered on each map (45.1\,arcmin for CBI1 and 28.2\,arcmin
  for CBI2).}
\label{figure:a1689_maps}
\end{figure*}

Observations of A1689 were carried out with CBI1 and CBI2 over the
periods 2004 May -- June and 2008 January -- May respectively, with
the pointing centre at $\rm{RA}(2000) = 13^{\rm{h}} 11^{\rm{m}}
29\fs5$ and $\rm{Dec.}~(2000) = -01\degr 20\arcmin 10\farcs0$.  For
both observations we adopted a similar observing strategy to that
described by \cite{Udomprasert:2004}, whereby any strong correlated
ground signal is subtracted out using reference fields separated by
8\,minutes in right ascension. This procedure increases the noise
level in the data by a factor of $\sqrt{2}$ in the case of a single
reference field, or $\sqrt{3/2}$ in the case of averaging over two
reference fields. Flagging and calibration of the visibility data were
performed using \textsc{CBICAL}, a specialist data reduction package
designed for use on CBI data. A total of 20,259 good visibilities were
collected by CBI1 and 13,295 by CBI2, where a visibility represents a
single 8-minute scan on each of the main and trail fields for each of
the 78 baselines and 10 frequency channels. This corresponds to a total
equivalent observing time (main plus trail fields) of 6.9 hours by
CBI1 and 4.5 hours by CBI2. The amplitude and phase were calibrated to
nightly observations of Jupiter and Tau A, or a suitable unresolved
source when neither primary source was available. The calibration is
ultimately tied to the measured brightness temperature of Jupiter at
33\,GHz, $T_{\rm{J}} = 146.6\pm0.75\,\rm{K}$ \citep[][]{Hill:2009},
and this introduces a 0.5 per cent calibration uncertainty in the
data. In addition to the absolute flux calibration, short observations
of secondary calibrators (including 3C 273, 3C 274, 3C 279, J1924-2292
and J2253+1610) were used to characterize any possible residual
pointing error in the experiment. These observations show that the
data are consistent with a residual pointing error of $0.5\,$ arcmin.

Full-resolution maps of both observations are shown in
Figure\,\ref{figure:a1689_maps}, and were deconvolved using the
\textsc{CLEAN} algorithm \citep{Hogbom:1974} implemented by the
\textsc {APCLN} task in \textsc{AIPS}\footnote{http://www.aips.nrao.edu/}. The decrement in
brightness due to the thermal SZ effect can clearly be seen at the
centres of both maps. The calibrated visibilities were gridded into
regularly spaced estimators using the
\textsc{MPIGRIDR}\footnote{http://www.cita.utoronto.ca/$\sim$myers/}
program, which implements the technique developed by
\cite{Myers:2003}. This significantly reduces the amount of data and
the size of the covariance matrices that need to be processed during
the model fitting.

\subsection{Additional sources of error}

\subsubsection{Intrinsic CMB anisotropy}

In addition to the instrumental thermal noise and calibration, a
significant source of uncertainty for CBI1 SZ observations is the
presence of the intrinsic CMB anisotropy \citep[]{Udomprasert:2004}.
The CMB contributions in the gridded visibilities are correlated and
therefore must be treated by the construction of a covariance
matrix. The intrinsic CMB covariances at the positions of the gridded
visibility data are constructed based on a model estimate of the CMB
power spectrum. \textsc{MPIGRIDR} is used to construct the covariance
matrix, with an input power spectrum generated using
\textsc{CMBFAST}\footnote{http://lambda.gsfc.nasa.gov/toolbox/tb\_cmbfast\_ov.cfm}
\citep{Seljak:1996}, assuming a flat $\Lambda$CDM cosmology with
$\Omega_{\rm{M}} = 0.3$, $\Omega_{\Lambda} = 0.7$ and $h = 0.7$. The
uncertainty due to the intrinsic CMB dominates on the largest scales
and so has the greatest effect on data from the CBI1 array, which has
shorter minimum baselines than the CBI2
array. Figures\,\ref{figure:a1689_results_1} and
\ref{figure:a1689_results_2} show the binned real visibility data as a
function of baseline and include the estimated uncertainty due to the
intrinsic CMB anisotropy.

\begin{table*}
  \centering
  \caption{Posterior estimates of the model parameter $\Delta T_{0}$
    and derived parameters $Y_{\rm{2500}}$, $Y^{\rm flat}_{\rm{2500}}$,
    $Y_{\rm{200}}$ and $Y^{\rm flat}_{\rm{200}}$ for A1689 from fitting
    to CBI1 and CBI2 data. The other model parameters are $\beta$,
    which is given a strong Gaussian prior of $\beta = 0.688 \pm
    0.013$, and $r_{\rm core}$, which is given either a strong
    Gaussian prior of $r_{\rm core} = 68.4 \pm 2.1$, or a flat
    uniform prior. $Y_{\rm{2500}}$ and $Y_{\rm{200}}$ are calculated by
    integrating the derived Comptonization parameter within projected
    radii of $r_{2500}$ (200\,arcsec) and $r_{200}$ (600\,arcsec)
    respectively using the strong prior on $r_{\rm core}$. $Y^{\rm
      flat}_{\rm{2500}}$ and $Y^{\rm flat}_{\rm{200}}$ are derived from
    model fitting to the data with a uniform prior on
    $r_{\rm{core}}$. $\Delta T_0$ is derived from the strong prior on
    $r_{\rm core}$ only. Error intervals represent 68\,per\,cent
    confidence. Priors on $r_{\rm{core}}$ and $\beta$ are from \citet{LaRoque:2006}.}
\begin{tabular}{lccc}
  \hline
  Parameter & CBI1 & CBI2 & Combined \\     
  \hline
  $\Delta T_{0}\,(\rm{mK})$ & $-1.09^{+0.24}_{-0.22}$ & $-1.23^{+0.24}_{-0.26}$ & $-1.20^{+0.18}_{-0.18}$ \\
  $Y_{\rm{2500}}\,(10^{-10}\,\rm{sr})$ & $1.79^{+0.36}_{-0.39}$ & $2.02^{+0.42}_{-0.40}$ & $1.95^{+0.33}_{-0.28}$ \\
  $Y^{\rm flat}_{\rm{2500}}\,(10^{-10}\,\rm{sr})$ & $1.12^{+0.56}_{-0.45}$ & $1.65^{+0.46}_{-0.46}$ & $1.56^{+0.34}_{-0.38}$ \\
  $Y_{\rm{200}}\,(10^{-10}\,\rm{sr})$ & $7.00^{+1.65}_{-1.45}$ & $8.12^{+1.71}_{-1.73}$ & $7.71^{+1.48}_{-1.16}$ \\
  $Y^{\rm flat}_{\rm{200}}\,(10^{-10}\,\rm{sr})$ & $3.65^{+2.40}_{-1.5}$ & $5.79^{+2.11}_{-1.88}$ & $5.25^{+1.68}_{-1.31}$ \\
  \hline
\end{tabular}
\label{table:a1689_results}
\end{table*}

\subsubsection{Point sources}

Bright point sources can contaminate the SZ signal, appearing as
positive sources if in the main field or negative if in a reference
field. Sources that are close to the field centres can have a
significant effect on the SZ decrement, especially since there is
relatively low attenuation from the primary beam. Spatial filtering of
the CBI1 and CBI2 data to angular scales smaller than 10\,arcmin
(corresponding to baselines longer than 300\,wavelengths) reveals no
significant point source flux density above the noise level. Two
sources were identified by \cite{Reese:2002} at 30\,GHz from BIMA and
OVRO observations of A1689. These sources are at positions
$(\rm{RA}(2000), \rm{Dec.}~(2000)) = (13^{\rm{h}} 11^{\rm{m}}
31\fs6, -01\degr 19\arcmin 33\farcs0)$ and $(13^{\rm{h}} 11^{\rm{m}}
30\fs1, -01\degr 20\arcmin 37\farcs0)$, and have integrated flux
densities of $1.33\pm0.10\,\rm{mJy}$ and $0.45\pm0.09\,\rm{mJy}$
respectively. The flux density contribution is unlikely to cause
significant contamination to the CBI data; however, we do subtract
these sources from the gridded visibility data. Any further residual
point source error is accounted for in the systematic error estimate
for model fitting.

\subsubsection{The Kinematic SZ effect}
The non-zero peculiar velocities of galaxy clusters cause a secondary
distortion in the CMB frequency spectrum known as the kinematic SZ
(KSZ) effect. \cite{Benson:2003} estimated the peculiar velocity of
A1689 to be $v_{\rm{pec}} = +170^{+805}_{-600}\pm750\,\rm{km\,s^{-1}}$
using the SuZIE II experiment (where the first quoted errors are
statistical and the second systematics). Clearly the uncertainties in
this measurement dominate, however we can estimate the uncertainty
introduced by the KSZ effect by assuming a typical line-of-sight
peculiar velocity of $\pm300\,\rm{km\,s^{-1}}$ \citep{Watkins:1997,
  Giovanelli:1998, Dale:1999, Colberg:2000}. At an observing frequency
of 31\,GHz the non-relativistic KSZ effect signal for a 10.5\,keV
cluster is $\pm$2.6\,per\,cent of the thermal SZ signal. Further
relativistic corrections to the KSZ signal have been calculated by
\cite{Nozawa:1998} and change the error by only 0.2--0.3\,per\,cent of
the thermal SZ signal for high temperature clusters.

\subsection{Model fitting}

We fit an analytical model to the visibility data in order to derive
properties of the cluster that can easily be compared with the
literature values. We assume that the observed thermal SZ effect has
circular symmetry and we model the change in brightness temperature
using the single isothermal $\beta$ model given by equation
\ref{eqn:isobeta}. Model fitting is done in visibility space where the
instrumental noise covariance matrix is diagonal (although radio sources and the CMB introduce off-diagonal elements to the covariance matrix). Implementation of
the model fitting is performed using \textsc{MultiNest}
\citep{Feroz:2009}, a powerful Bayesian optimizer that uses the Nested
Sampling method \citep{Skilling:2004}. This program returns a weighted
sampled posterior probability distribution for each of the model
parameters, which can then be marginalized over in order to obtain
estimates of the derived cluster properties.  We also introduce a
calibration error as a nuisance parameter with a Gaussian prior that
accounts for a total systematic error of 5\,per cent (one sigma), and
which is later marginalized over.

\begin{figure*}
\centering
\includegraphics[width = 0.3\textwidth]{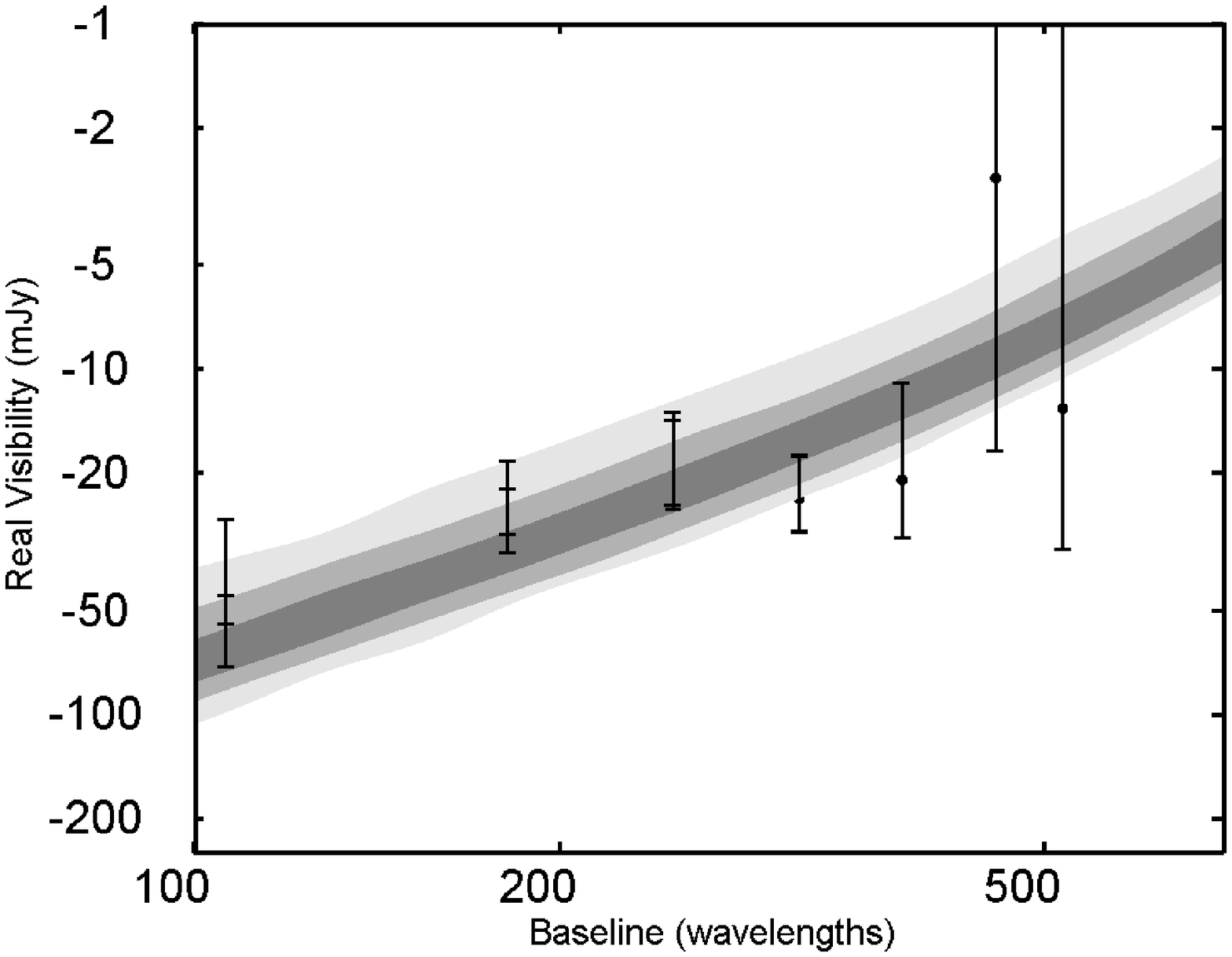}
\includegraphics[width = 0.29\textwidth]{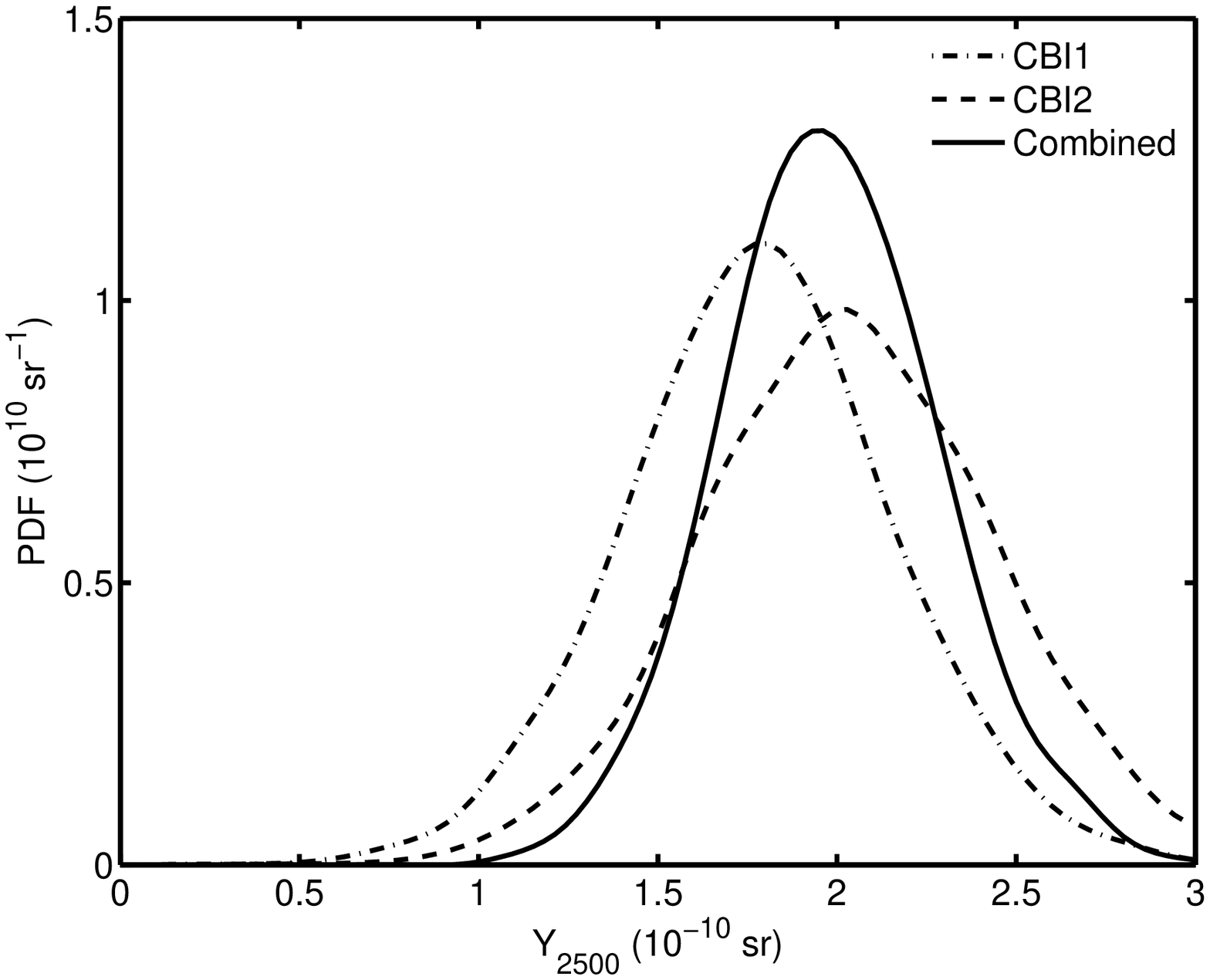}
\includegraphics[width = 0.3\textwidth]{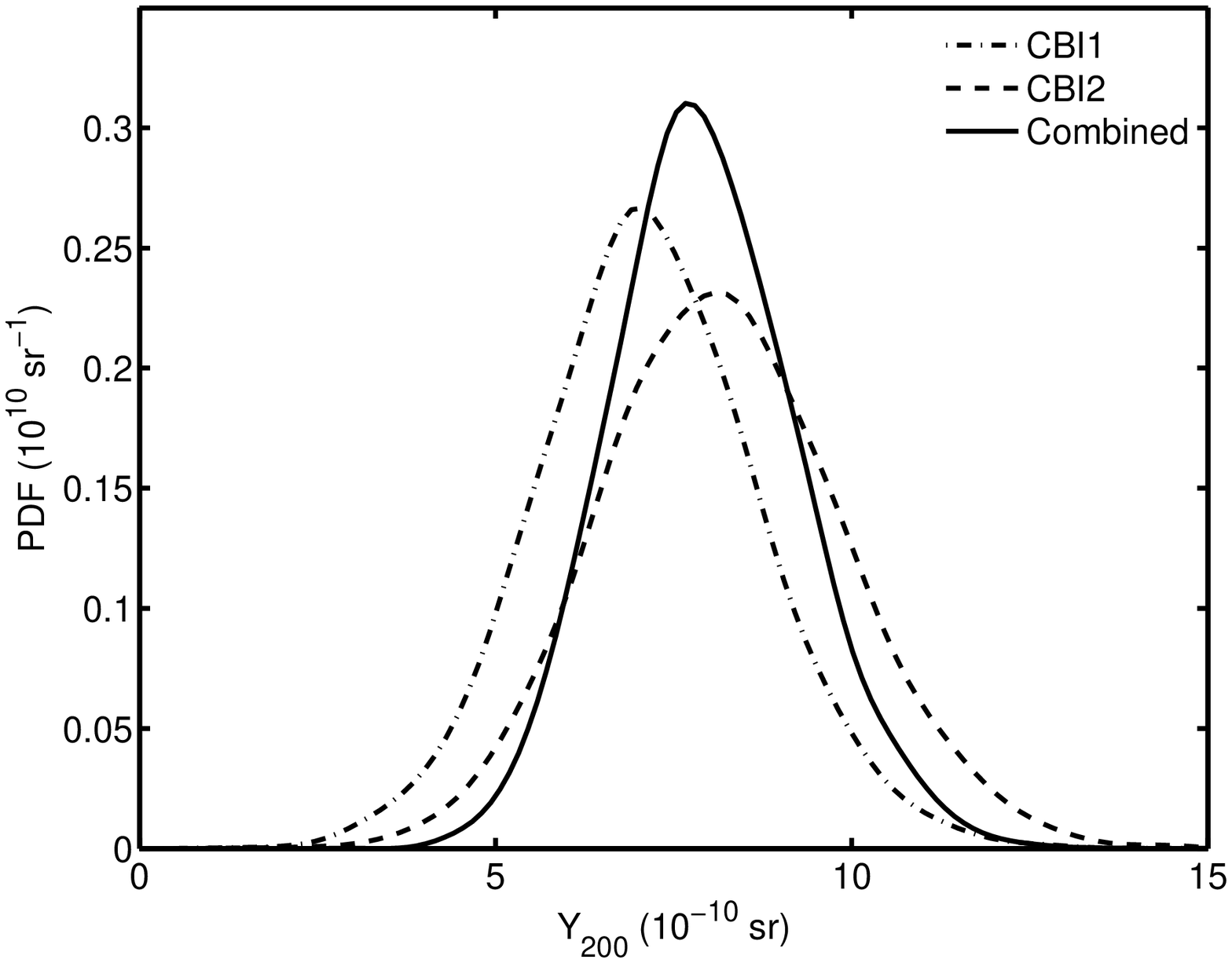}
\caption{Results from the fit of an isothermal beta model to the
  combined CBI1 and CBI2 data, using the priors on $\beta$ and
  $r_{\rm{core}}$ from \citet{LaRoque:2006} ($\beta = 0.688 \pm .013$,
  $r_{\rm core} = 48.4 \pm 2.1$~arcsec). \emph{Left:} The real part
  of the visibility for the combined CBI data set as a function of
  baseline distance. The short errorbars represent the $1\,\sigma$
  noise from the variance in the data, and the long errorbars
  represent the total statistical uncertainty including the intrinsic
  CMB anisotropy. The grey scale represents the 68.3, 95.4 and
  99.7\,per\,cent confidence intervals of the fitted
  model. \emph{Middle and Right:} The estimated posterior probability
  distributions for the integrated Comptonization parameters
  $Y_{\rm{2500}}$ and $Y_{\rm{200}}$.}
\label{figure:a1689_results_1}
\end{figure*}

\begin{figure*}
\centering
\includegraphics[width = 0.3\textwidth]{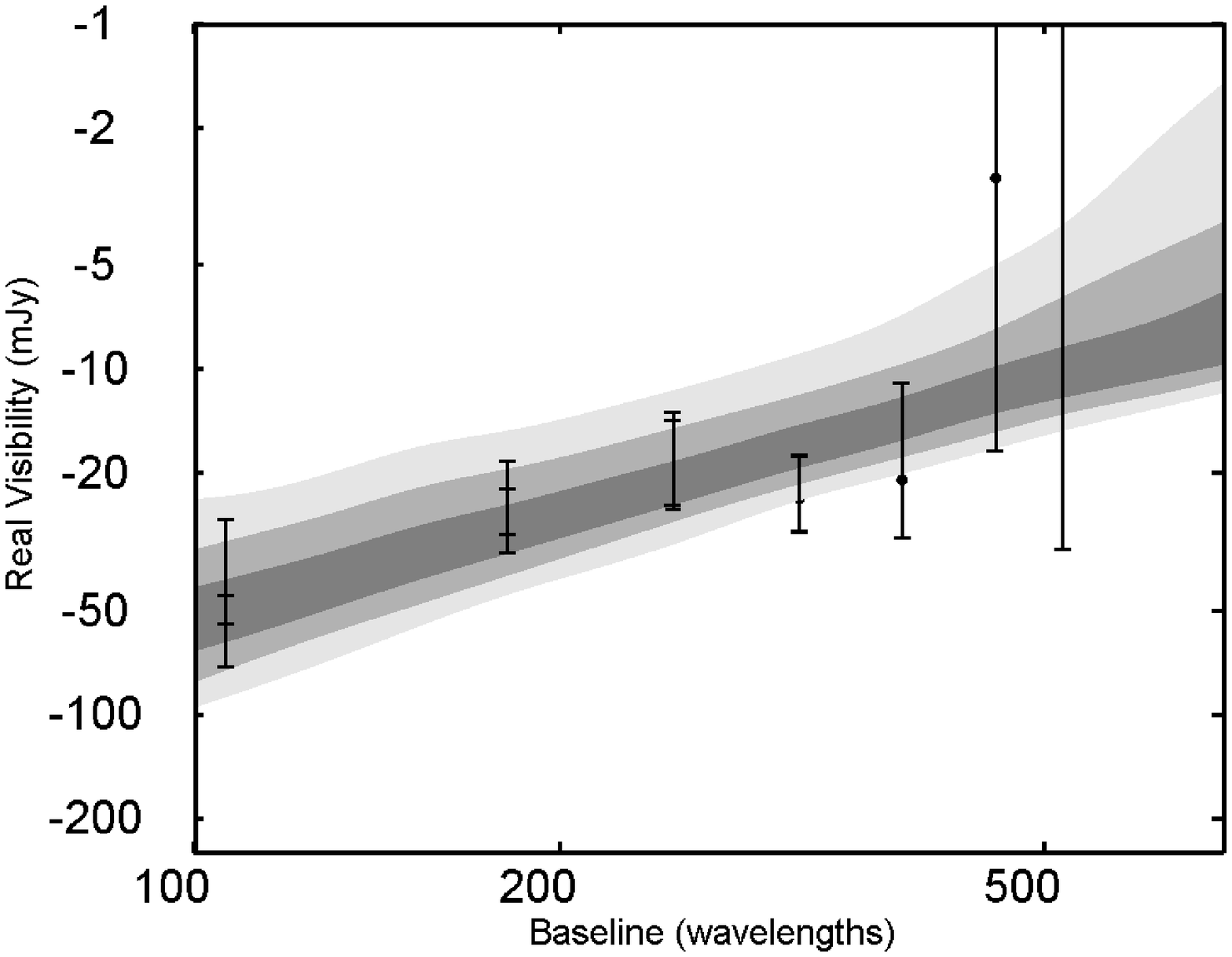}
\includegraphics[width = 0.29\textwidth]{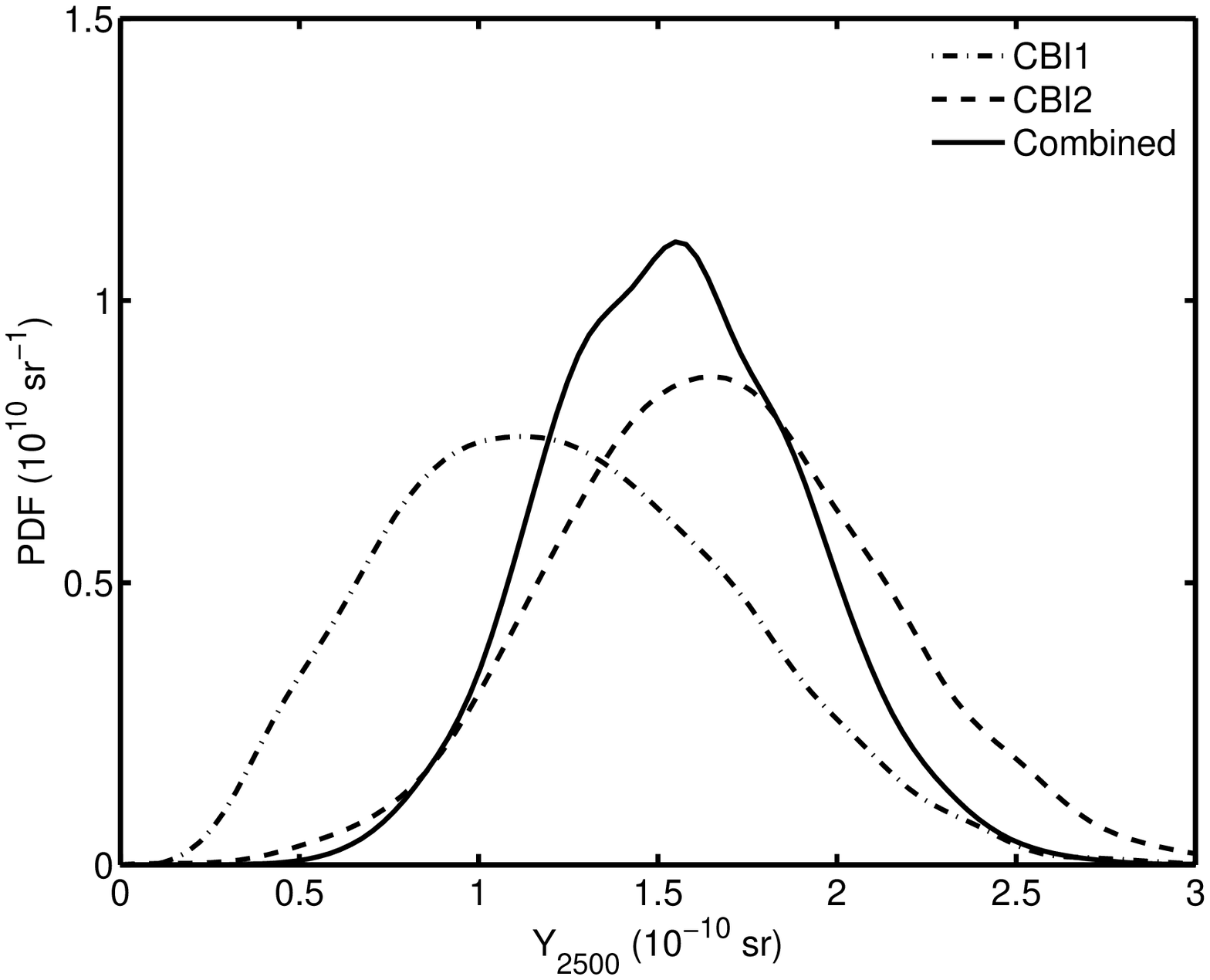}
\includegraphics[width = 0.3\textwidth]{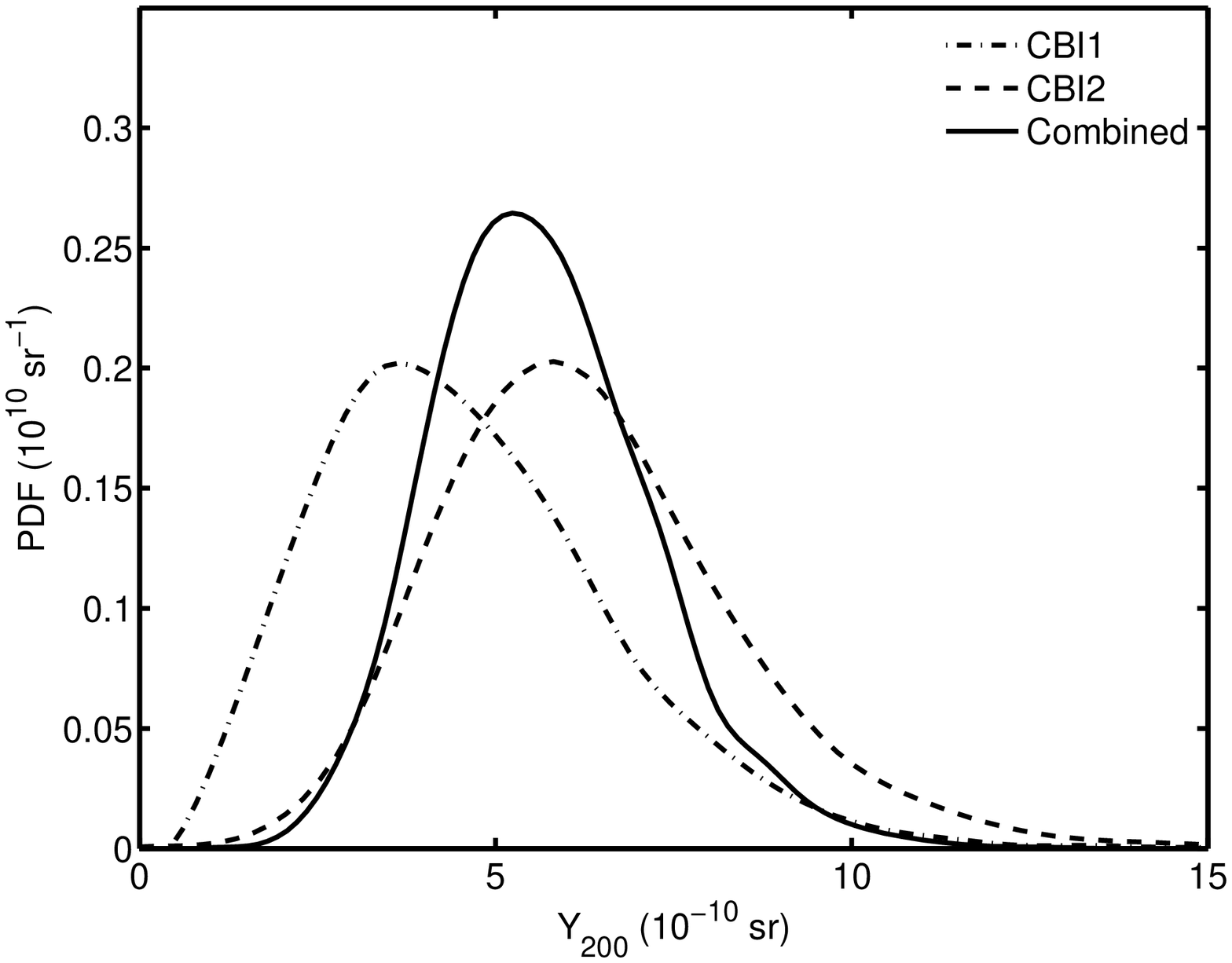}\\
\includegraphics[trim=0mm 0mm 9mm 0mm,clip,width = 0.31\textwidth]{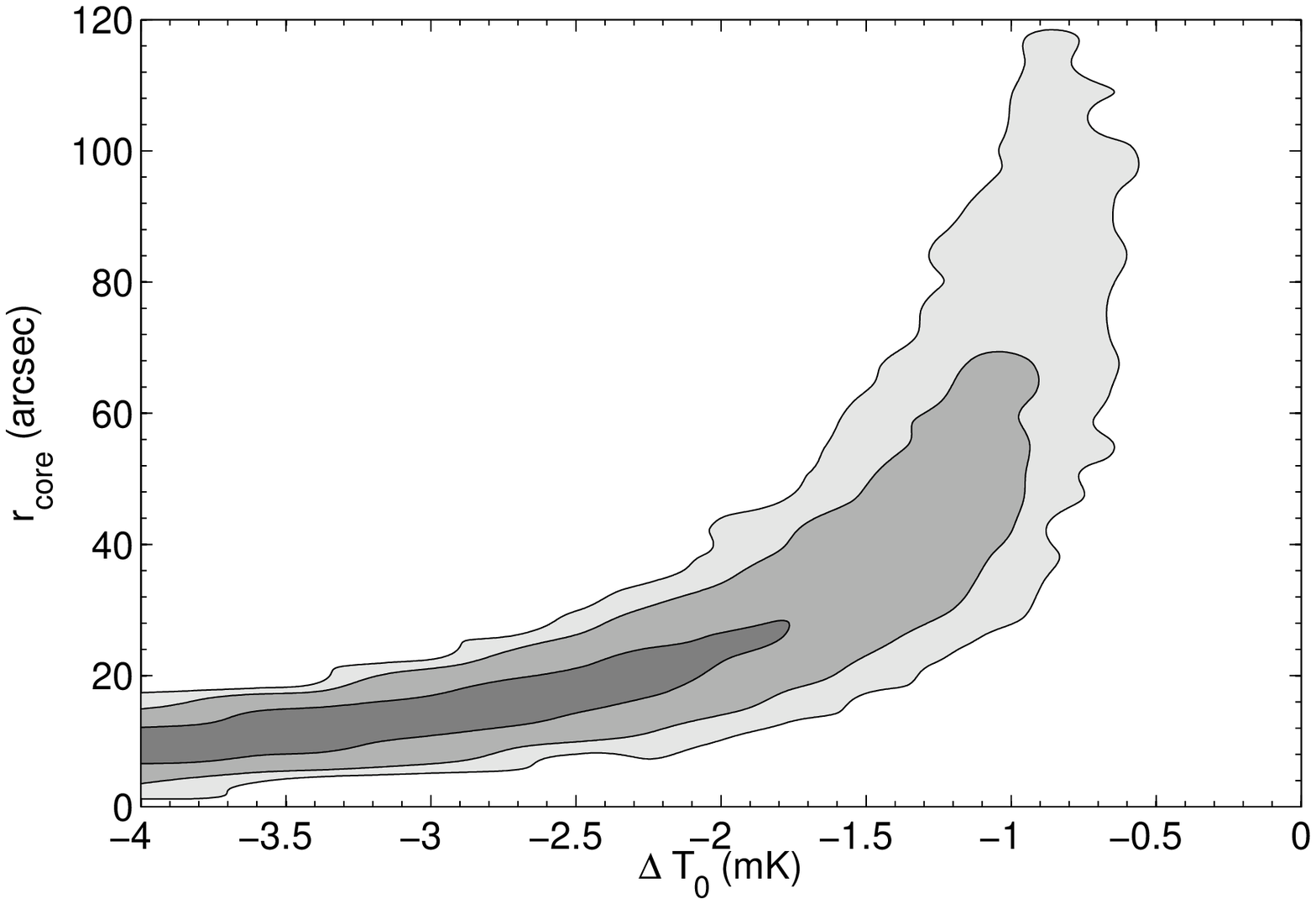}
\caption{(Top) As Figure \ref{figure:a1689_results_1} but with a
  uniform prior on $r_{\rm core}$, showing that reasonable
  constraints can be placed on $Y$ even without detailed prior
  knowledge of the cluster core size. (Bottom) The joint constraint on
  $\Delta T_0$ and $r_{\rm core}$, showing that they are individually
  poorly constrained. }
\label{figure:a1689_results_2}
\end{figure*}

\subsection{Results}

Table\,\ref{table:a1689_results} shows the priors and estimates of
model parameters from fitting separately to CBI1 and CBI2 data, and
then jointly to both. The $(u,v)$ coverage of the CBI arrays means
that $r_{\rm{core}}$ and $\beta$ are not individually well constrained
by the data and so we apply Gaussian priors to these parameters based
on the values measured by \cite{LaRoque:2006} from a combined X-ray
and SZ analysis. The estimated values of the central SZ decrement for
the CBI1 and CBI2 arrays are consistent within the errors. Although
the error bars on the parameters (which are derived from the posterior
probability distributions) are similar for CBI1 and CBI2, the joint
fit is dominated by the CBI2 data. This is because the CBI1 data have
significant off-diagonal terms in the covariance matrix due to the
intrinsic CMB fluctuations, which do not integrate down with the
addition of more data. Our joint fitting to both data sets fully takes
this effect into account, but it means one cannot simply take the
weighted mean of the individual CBI1 and CBI2 parameter estimates.

From the model we can calculate the total integrated $y$-parameter for a
given projected aperture. This quantity is a measure of the cluster's
total thermal energy contained within the radius of integration, and
is given by
\begin{equation}
  Y \equiv \int_{\rm{A}}{y\,\rm{d}\Omega} =  2\pi\int_{\rm{r}}{y r\,\rm{d}r},
\end{equation}
where $r$ is the projected angular radius from the centroid. $Y$ is
the most interesting observable parameter in terms of relating
measured SZ signals to the intrinsic cluster properties such as mass.
In order to compare with values given in the literature we calculate
$Y_{\rm{2500}}$ within a radius of 200\,arcsec , which corresponds
roughly to an overdensity radius of $r_{2500}$. \cite{Bonamente:2008}
used observations of A1689 with the BIMA and OVRO arrays to obtain an
estimate of $Y_{2500} = 1.88^{+0.49}_{-0.38}\,\times\,10^{-10}\,\rm
sr$ with $r_{2500} = 196\pm8\,\rm{arcsec}$, and \cite{Liao:2010} used
AMIBA observations to obtain an estimate of $Y_{2500} =
3.1^{+1.3}_{-1.3}\,\times\,10^{-10}\,\rm sr$ with $r_{2500} =
215^{+16}_{-19}\,\rm{arcsec}$. Figure\,\ref{figure:a1689_results_1}
shows the posterior probability distribution for estimates of $Y$ from
CBI1 and CBI2 observations. Our estimate of $Y_{\rm{2500}} =
1.95^{+0.33}_{-0.28} \times 10^{-10}\,\rm sr$ is consistent with these
results, within the errors.

We also calculate $Y$ out to a larger projected radius of 600\,arcsec,
which is approximately the value of $r_{200}$ quoted by
\cite{LaRoque:2006}, in order to demonstrate that the CBI1 and CBI2
arrays are capable of measuring the thermal SZ effect integrated out
to large physical radii. The value of $Y$ at $r_{200}$ was calculated
using the strong prior on $r_{\rm core}$ used above, giving
$Y_{\rm{200}} = 7.71^{+1.48}_{-1.16} \times 10^{-10}\,\rm sr$. This
significantly larger value than $Y_{2500}$ reflects the fact that a
large fraction of the integrated cluster pressure is outside the core
region probed by higher resolution experiments. The fit to the CBI
visibility data and the posterior probability distributions for
$Y_{\rm{200}}$ and $Y_{\rm{2500}}$ are shown in Figure
\ref{figure:a1689_results_1}.

We have used prior values of $\beta$ and $r_{\rm core}$ derived from
X-ray measurements to estimate $\Delta T_0$, as our SZ data by
themselves do not put strong constraints on these parameters. However,
it is not necessary to accurately constrain all the parameters of the
$\beta$ model in order to make a good measurement of $Y$. We therefore also
calculate $Y^{\rm flat}$ using a flat prior on $r_{\rm core}$, i.e.,
assuming we have no knowledge of this parameter, while maintaining the
strong prior on $\beta$, which typically does not vary significantly
from a value of $\beta \simeq 2/3$ between different clusters. Figure
\ref{figure:a1689_results_2} shows the fit to the CBI visibility data,
the probability distributions for $Y^{\rm flat}_{\rm{200}}$ and $Y^{\rm
  flat}_{\rm{2500}}$, and the likelihood contours for the free
parameters $r_{\rm core}$ and $\Delta T_0$. Although $r_{\rm core}$
and $\Delta T_0$ are individually very poorly constrained, the fits for
$Y^{\rm flat}$ are only slightly worse than when $r_{\rm core}$ is
strongly constrained. The estimates of $Y^{\rm flat}$ are also
significantly improved when fitting to the combined CBI1 and CBI2 data
sets. This is to be expected since $Y$ is proportional to the SZ total
flux density (i.e., the zero-spacing visibility), which is measured
almost directly by the CBI short baselines.

\section{Conclusions}
We have described an upgrade to the Cosmic Background Imager in which the original thirteen 0.9~m antennas were replaced with new 1.4~m antennas. The upgrade was achieved by using inexpensive, commercial off-the-shelf antennas with custom-made secondary
mirrors. The fabrication techniques were designed to be cheap and easy
to reproduce and could be used as a cost-effective method for
producing a large number of antennas in this size range. The new
antennas have been demonstrated to perform as specified, and using SZ
observations of the cluster A1689 we have shown that the
inter-calibration of CBI2 and CBI1 is good, that the upgraded array
met its design sensitivity, and that CBI2 and combined CBI1 plus CBI2
data can be used to constrain the integrated SZ Comptonization
parameter out to large radii ($r_{200}$) both with and without
prior information on the cluster size.

\section*{Acknowledgments} 

ACT thanks the Royal Society for support via a Dorothy Hodgkin
Research Fellowship and a Small Research Grant. ACT also acknowledges
an STFC Post-doctoral Research Fellowship. The Royal Society have also
supported CBI2 via an International Joint Project Grant awarded to
MEJ.  JRA acknowledges support by a studentship from the Science and
Technology Facilities Council and a Super Science Fellowship from the Australian Research Council. LB and RB acknowledge support from
Center of Excellence in Astrophysics and Associated Technologies (PFB
06), and RB also acknowledges support from ALMA-Conicyt 31080022 and
31070015. CD acknowledges an STFC advanced fellowship and ERC IRG
grant under the FP7.  We thank the contributing institutions and
investigators of the Strategic Alliance for the Implementation of New
Technologies (SAINT) for vital financial support of the Chajnantor
Observatory. We also thank the Kavli Operating Institute, Barbara and
Stanley Rawn Jr., Maxine and Ronald Linde, Cecil and Sally Drinkward,
and Rochus Vogt. We thank Matthias Tecza for help with the ZEMAX
analysis. Finally, we thank the staff of the Chajnantor Observatory,
particularly Crist\'{o}bal Achermann, Nolberto Oyarce, Jos\'{e} Cortes
and Wilson Araya, for expert and untiring support of the operations of
CBI.

\bibliography{cbi_instr}

\label{lastpage}

\end{document}

%% file: rgb.tex
\definecolor{AliceBlue}{rgb}{0.94,0.97,1.00}
\definecolor{AntiqueWhite1}{rgb}{1.00,0.94,0.86}
\definecolor{AntiqueWhite2}{rgb}{0.93,0.87,0.80}
\definecolor{AntiqueWhite3}{rgb}{0.80,0.75,0.69}
\definecolor{AntiqueWhite4}{rgb}{0.55,0.51,0.47}
\definecolor{AntiqueWhite}{rgb}{0.98,0.92,0.84}
\definecolor{BlanchedAlmond}{rgb}{1.00,0.92,0.80}
\definecolor{BlueViolet}{rgb}{0.54,0.17,0.89}
\definecolor{CadetBlue1}{rgb}{0.60,0.96,1.00}
\definecolor{CadetBlue2}{rgb}{0.56,0.90,0.93}
\definecolor{CadetBlue3}{rgb}{0.48,0.77,0.80}
\definecolor{CadetBlue4}{rgb}{0.33,0.53,0.55}
\definecolor{CadetBlue}{rgb}{0.37,0.62,0.63}
\definecolor{CornflowerBlue}{rgb}{0.39,0.58,0.93}
\definecolor{DarkBlue}{rgb}{0.00,0.00,0.55}
\definecolor{DarkCyan}{rgb}{0.00,0.55,0.55}
\definecolor{DarkGoldenrod1}{rgb}{1.00,0.73,0.06}
\definecolor{DarkGoldenrod2}{rgb}{0.93,0.68,0.05}
\definecolor{DarkGoldenrod3}{rgb}{0.80,0.58,0.05}
\definecolor{DarkGoldenrod4}{rgb}{0.55,0.40,0.03}
\definecolor{DarkGoldenrod}{rgb}{0.72,0.53,0.04}
\definecolor{DarkGray}{rgb}{0.66,0.66,0.66}
\definecolor{DarkGreen}{rgb}{0.00,0.39,0.00}
\definecolor{DarkGrey}{rgb}{0.66,0.66,0.66}
\definecolor{DarkKhaki}{rgb}{0.74,0.72,0.42}
\definecolor{DarkMagenta}{rgb}{0.55,0.00,0.55}
\definecolor{DarkOliveGreen1}{rgb}{0.79,1.00,0.44}
\definecolor{DarkOliveGreen2}{rgb}{0.74,0.93,0.41}
\definecolor{DarkOliveGreen3}{rgb}{0.64,0.80,0.35}
\definecolor{DarkOliveGreen4}{rgb}{0.43,0.55,0.24}
\definecolor{DarkOliveGreen}{rgb}{0.33,0.42,0.18}
\definecolor{DarkOrange1}{rgb}{1.00,0.50,0.00}
\definecolor{DarkOrange2}{rgb}{0.93,0.46,0.00}
\definecolor{DarkOrange3}{rgb}{0.80,0.40,0.00}
\definecolor{DarkOrange4}{rgb}{0.55,0.27,0.00}
\definecolor{DarkOrange}{rgb}{1.00,0.55,0.00}
\definecolor{DarkOrchid1}{rgb}{0.75,0.24,1.00}
\definecolor{DarkOrchid2}{rgb}{0.70,0.23,0.93}
\definecolor{DarkOrchid3}{rgb}{0.60,0.20,0.80}
\definecolor{DarkOrchid4}{rgb}{0.41,0.13,0.55}
\definecolor{DarkOrchid}{rgb}{0.60,0.20,0.80}
\definecolor{DarkRed}{rgb}{0.55,0.00,0.00}
\definecolor{DarkSalmon}{rgb}{0.91,0.59,0.48}
\definecolor{DarkSeaGreen1}{rgb}{0.76,1.00,0.76}
\definecolor{DarkSeaGreen2}{rgb}{0.71,0.93,0.71}
\definecolor{DarkSeaGreen3}{rgb}{0.61,0.80,0.61}
\definecolor{DarkSeaGreen4}{rgb}{0.41,0.55,0.41}
\definecolor{DarkSeaGreen}{rgb}{0.56,0.74,0.56}
\definecolor{DarkSlateBlue}{rgb}{0.28,0.24,0.55}
\definecolor{DarkSlateGray1}{rgb}{0.59,1.00,1.00}
\definecolor{DarkSlateGray2}{rgb}{0.55,0.93,0.93}
\definecolor{DarkSlateGray3}{rgb}{0.47,0.80,0.80}
\definecolor{DarkSlateGray4}{rgb}{0.32,0.55,0.55}
\definecolor{DarkSlateGray}{rgb}{0.18,0.31,0.31}
\definecolor{DarkSlateGrey}{rgb}{0.18,0.31,0.31}
\definecolor{DarkTurquoise}{rgb}{0.00,0.81,0.82}
\definecolor{DarkViolet}{rgb}{0.58,0.00,0.83}
\definecolor{DeepPink1}{rgb}{1.00,0.08,0.58}
\definecolor{DeepPink2}{rgb}{0.93,0.07,0.54}
\definecolor{DeepPink3}{rgb}{0.80,0.06,0.46}
\definecolor{DeepPink4}{rgb}{0.55,0.04,0.31}
\definecolor{DeepPink}{rgb}{1.00,0.08,0.58}
\definecolor{DeepSkyBlue1}{rgb}{0.00,0.75,1.00}
\definecolor{DeepSkyBlue2}{rgb}{0.00,0.70,0.93}
\definecolor{DeepSkyBlue3}{rgb}{0.00,0.60,0.80}
\definecolor{DeepSkyBlue4}{rgb}{0.00,0.41,0.55}
\definecolor{DeepSkyBlue}{rgb}{0.00,0.75,1.00}
\definecolor{DimGray}{rgb}{0.41,0.41,0.41}
\definecolor{DimGrey}{rgb}{0.41,0.41,0.41}
\definecolor{DodgerBlue1}{rgb}{0.12,0.56,1.00}
\definecolor{DodgerBlue2}{rgb}{0.11,0.53,0.93}
\definecolor{DodgerBlue3}{rgb}{0.09,0.45,0.80}
\definecolor{DodgerBlue4}{rgb}{0.06,0.31,0.55}
\definecolor{DodgerBlue}{rgb}{0.12,0.56,1.00}
\definecolor{FloralWhite}{rgb}{1.00,0.98,0.94}
\definecolor{ForestGreen}{rgb}{0.13,0.55,0.13}
\definecolor{GhostWhite}{rgb}{0.97,0.97,1.00}
\definecolor{GreenYellow}{rgb}{0.68,1.00,0.18}
\definecolor{HotPink1}{rgb}{1.00,0.43,0.71}
\definecolor{HotPink2}{rgb}{0.93,0.42,0.65}
\definecolor{HotPink3}{rgb}{0.80,0.38,0.56}
\definecolor{HotPink4}{rgb}{0.55,0.23,0.38}
\definecolor{HotPink}{rgb}{1.00,0.41,0.71}
\definecolor{IndianRed1}{rgb}{1.00,0.42,0.42}
\definecolor{IndianRed2}{rgb}{0.93,0.39,0.39}
\definecolor{IndianRed3}{rgb}{0.80,0.33,0.33}
\definecolor{IndianRed4}{rgb}{0.55,0.23,0.23}
\definecolor{IndianRed}{rgb}{0.80,0.36,0.36}
\definecolor{LavenderBlush1}{rgb}{1.00,0.94,0.96}
\definecolor{LavenderBlush2}{rgb}{0.93,0.88,0.90}
\definecolor{LavenderBlush3}{rgb}{0.80,0.76,0.77}
\definecolor{LavenderBlush4}{rgb}{0.55,0.51,0.53}
\definecolor{LavenderBlush}{rgb}{1.00,0.94,0.96}
\definecolor{LawnGreen}{rgb}{0.49,0.99,0.00}
\definecolor{LemonChiffon1}{rgb}{1.00,0.98,0.80}
\definecolor{LemonChiffon2}{rgb}{0.93,0.91,0.75}
\definecolor{LemonChiffon3}{rgb}{0.80,0.79,0.65}
\definecolor{LemonChiffon4}{rgb}{0.55,0.54,0.44}
\definecolor{LemonChiffon}{rgb}{1.00,0.98,0.80}
\definecolor{LightBlue1}{rgb}{0.75,0.94,1.00}
\definecolor{LightBlue2}{rgb}{0.70,0.87,0.93}
\definecolor{LightBlue3}{rgb}{0.60,0.75,0.80}
\definecolor{LightBlue4}{rgb}{0.41,0.51,0.55}
\definecolor{LightBlue}{rgb}{0.68,0.85,0.90}
\definecolor{LightCoral}{rgb}{0.94,0.50,0.50}
\definecolor{LightCyan1}{rgb}{0.88,1.00,1.00}
\definecolor{LightCyan2}{rgb}{0.82,0.93,0.93}
\definecolor{LightCyan3}{rgb}{0.71,0.80,0.80}
\definecolor{LightCyan4}{rgb}{0.48,0.55,0.55}
\definecolor{LightCyan}{rgb}{0.88,1.00,1.00}
\definecolor{LightGoldenrod1}{rgb}{1.00,0.93,0.55}
\definecolor{LightGoldenrod2}{rgb}{0.93,0.86,0.51}
\definecolor{LightGoldenrod3}{rgb}{0.80,0.75,0.44}
\definecolor{LightGoldenrod4}{rgb}{0.55,0.51,0.30}
\definecolor{LightGoldenrodYellow}{rgb}{0.98,0.98,0.82}
\definecolor{LightGoldenrod}{rgb}{0.93,0.87,0.51}
\definecolor{LightGray}{rgb}{0.83,0.83,0.83}
\definecolor{LightGreen}{rgb}{0.56,0.93,0.56}
\definecolor{LightGrey}{rgb}{0.83,0.83,0.83}
\definecolor{LightPink1}{rgb}{1.00,0.68,0.73}
\definecolor{LightPink2}{rgb}{0.93,0.64,0.68}
\definecolor{LightPink3}{rgb}{0.80,0.55,0.58}
\definecolor{LightPink4}{rgb}{0.55,0.37,0.40}
\definecolor{LightPink}{rgb}{1.00,0.71,0.76}
\definecolor{LightSalmon1}{rgb}{1.00,0.63,0.48}
\definecolor{LightSalmon2}{rgb}{0.93,0.58,0.45}
\definecolor{LightSalmon3}{rgb}{0.80,0.51,0.38}
\definecolor{LightSalmon4}{rgb}{0.55,0.34,0.26}
\definecolor{LightSalmon}{rgb}{1.00,0.63,0.48}
\definecolor{LightSeaGreen}{rgb}{0.13,0.70,0.67}
\definecolor{LightSkyBlue1}{rgb}{0.69,0.89,1.00}
\definecolor{LightSkyBlue2}{rgb}{0.64,0.83,0.93}
\definecolor{LightSkyBlue3}{rgb}{0.55,0.71,0.80}
\definecolor{LightSkyBlue4}{rgb}{0.38,0.48,0.55}
\definecolor{LightSkyBlue}{rgb}{0.53,0.81,0.98}
\definecolor{LightSlateBlue}{rgb}{0.52,0.44,1.00}
\definecolor{LightSlateGray}{rgb}{0.47,0.53,0.60}
\definecolor{LightSlateGrey}{rgb}{0.47,0.53,0.60}
\definecolor{LightSteelBlue1}{rgb}{0.79,0.88,1.00}
\definecolor{LightSteelBlue2}{rgb}{0.74,0.82,0.93}
\definecolor{LightSteelBlue3}{rgb}{0.64,0.71,0.80}
\definecolor{LightSteelBlue4}{rgb}{0.43,0.48,0.55}
\definecolor{LightSteelBlue}{rgb}{0.69,0.77,0.87}
\definecolor{LightYellow1}{rgb}{1.00,1.00,0.88}
\definecolor{LightYellow2}{rgb}{0.93,0.93,0.82}
\definecolor{LightYellow3}{rgb}{0.80,0.80,0.71}
\definecolor{LightYellow4}{rgb}{0.55,0.55,0.48}
\definecolor{LightYellow}{rgb}{1.00,1.00,0.88}
\definecolor{LimeGreen}{rgb}{0.20,0.80,0.20}
\definecolor{MediumAquamarine}{rgb}{0.40,0.80,0.67}
\definecolor{MediumBlue}{rgb}{0.00,0.00,0.80}
\definecolor{MediumOrchid1}{rgb}{0.88,0.40,1.00}
\definecolor{MediumOrchid2}{rgb}{0.82,0.37,0.93}
\definecolor{MediumOrchid3}{rgb}{0.71,0.32,0.80}
\definecolor{MediumOrchid4}{rgb}{0.48,0.22,0.55}
\definecolor{MediumOrchid}{rgb}{0.73,0.33,0.83}
\definecolor{MediumPurple1}{rgb}{0.67,0.51,1.00}
\definecolor{MediumPurple2}{rgb}{0.62,0.47,0.93}
\definecolor{MediumPurple3}{rgb}{0.54,0.41,0.80}
\definecolor{MediumPurple4}{rgb}{0.36,0.28,0.55}
\definecolor{MediumPurple}{rgb}{0.58,0.44,0.86}
\definecolor{MediumSeaGreen}{rgb}{0.24,0.70,0.44}
\definecolor{MediumSlateBlue}{rgb}{0.48,0.41,0.93}
\definecolor{MediumSpringGreen}{rgb}{0.00,0.98,0.60}
\definecolor{MediumTurquoise}{rgb}{0.28,0.82,0.80}
\definecolor{MediumVioletRed}{rgb}{0.78,0.08,0.52}
\definecolor{MidnightBlue}{rgb}{0.10,0.10,0.44}
\definecolor{MintCream}{rgb}{0.96,1.00,0.98}
\definecolor{MistyRose1}{rgb}{1.00,0.89,0.88}
\definecolor{MistyRose2}{rgb}{0.93,0.84,0.82}
\definecolor{MistyRose3}{rgb}{0.80,0.72,0.71}
\definecolor{MistyRose4}{rgb}{0.55,0.49,0.48}
\definecolor{MistyRose}{rgb}{1.00,0.89,0.88}
\definecolor{NavajoWhite1}{rgb}{1.00,0.87,0.68}
\definecolor{NavajoWhite2}{rgb}{0.93,0.81,0.63}
\definecolor{NavajoWhite3}{rgb}{0.80,0.70,0.55}
\definecolor{NavajoWhite4}{rgb}{0.55,0.47,0.37}
\definecolor{NavajoWhite}{rgb}{1.00,0.87,0.68}
\definecolor{NavyBlue}{rgb}{0.00,0.00,0.50}
\definecolor{OldLace}{rgb}{0.99,0.96,0.90}
\definecolor{OliveDrab1}{rgb}{0.75,1.00,0.24}
\definecolor{OliveDrab2}{rgb}{0.70,0.93,0.23}
\definecolor{OliveDrab3}{rgb}{0.60,0.80,0.20}
\definecolor{OliveDrab4}{rgb}{0.41,0.55,0.13}
\definecolor{OliveDrab}{rgb}{0.42,0.56,0.14}
\definecolor{OrangeRed1}{rgb}{1.00,0.27,0.00}
\definecolor{OrangeRed2}{rgb}{0.93,0.25,0.00}
\definecolor{OrangeRed3}{rgb}{0.80,0.22,0.00}
\definecolor{OrangeRed4}{rgb}{0.55,0.15,0.00}
\definecolor{OrangeRed}{rgb}{1.00,0.27,0.00}
\definecolor{PaleGoldenrod}{rgb}{0.93,0.91,0.67}
\definecolor{PaleGreen1}{rgb}{0.60,1.00,0.60}
\definecolor{PaleGreen2}{rgb}{0.56,0.93,0.56}
\definecolor{PaleGreen3}{rgb}{0.49,0.80,0.49}
\definecolor{PaleGreen4}{rgb}{0.33,0.55,0.33}
\definecolor{PaleGreen}{rgb}{0.60,0.98,0.60}
\definecolor{PaleTurquoise1}{rgb}{0.73,1.00,1.00}
\definecolor{PaleTurquoise2}{rgb}{0.68,0.93,0.93}
\definecolor{PaleTurquoise3}{rgb}{0.59,0.80,0.80}
\definecolor{PaleTurquoise4}{rgb}{0.40,0.55,0.55}
\definecolor{PaleTurquoise}{rgb}{0.69,0.93,0.93}
\definecolor{PaleVioletRed1}{rgb}{1.00,0.51,0.67}
\definecolor{PaleVioletRed2}{rgb}{0.93,0.47,0.62}
\definecolor{PaleVioletRed3}{rgb}{0.80,0.41,0.54}
\definecolor{PaleVioletRed4}{rgb}{0.55,0.28,0.36}
\definecolor{PaleVioletRed}{rgb}{0.86,0.44,0.58}
\definecolor{PapayaWhip}{rgb}{1.00,0.94,0.84}
\definecolor{PeachPuff1}{rgb}{1.00,0.85,0.73}
\definecolor{PeachPuff2}{rgb}{0.93,0.80,0.68}
\definecolor{PeachPuff3}{rgb}{0.80,0.69,0.58}
\definecolor{PeachPuff4}{rgb}{0.55,0.47,0.40}
\definecolor{PeachPuff}{rgb}{1.00,0.85,0.73}
\definecolor{PowderBlue}{rgb}{0.69,0.88,0.90}
\definecolor{RosyBrown1}{rgb}{1.00,0.76,0.76}
\definecolor{RosyBrown2}{rgb}{0.93,0.71,0.71}
\definecolor{RosyBrown3}{rgb}{0.80,0.61,0.61}
\definecolor{RosyBrown4}{rgb}{0.55,0.41,0.41}
\definecolor{RosyBrown}{rgb}{0.74,0.56,0.56}
\definecolor{RoyalBlue1}{rgb}{0.28,0.46,1.00}
\definecolor{RoyalBlue2}{rgb}{0.26,0.43,0.93}
\definecolor{RoyalBlue3}{rgb}{0.23,0.37,0.80}
\definecolor{RoyalBlue4}{rgb}{0.15,0.25,0.55}
\definecolor{RoyalBlue}{rgb}{0.25,0.41,0.88}
\definecolor{SaddleBrown}{rgb}{0.55,0.27,0.07}
\definecolor{SandyBrown}{rgb}{0.96,0.64,0.38}
\definecolor{SeaGreen1}{rgb}{0.33,1.00,0.62}
\definecolor{SeaGreen2}{rgb}{0.31,0.93,0.58}
\definecolor{SeaGreen3}{rgb}{0.26,0.80,0.50}
\definecolor{SeaGreen4}{rgb}{0.18,0.55,0.34}
\definecolor{SeaGreen}{rgb}{0.18,0.55,0.34}
\definecolor{SkyBlue1}{rgb}{0.53,0.81,1.00}
\definecolor{SkyBlue2}{rgb}{0.49,0.75,0.93}
\definecolor{SkyBlue3}{rgb}{0.42,0.65,0.80}
\definecolor{SkyBlue4}{rgb}{0.29,0.44,0.55}
\definecolor{SkyBlue}{rgb}{0.53,0.81,0.92}
\definecolor{SlateBlue1}{rgb}{0.51,0.44,1.00}
\definecolor{SlateBlue2}{rgb}{0.48,0.40,0.93}
\definecolor{SlateBlue3}{rgb}{0.41,0.35,0.80}
\definecolor{SlateBlue4}{rgb}{0.28,0.24,0.55}
\definecolor{SlateBlue}{rgb}{0.42,0.35,0.80}
\definecolor{SlateGray1}{rgb}{0.78,0.89,1.00}
\definecolor{SlateGray2}{rgb}{0.73,0.83,0.93}
\definecolor{SlateGray3}{rgb}{0.62,0.71,0.80}
\definecolor{SlateGray4}{rgb}{0.42,0.48,0.55}
\definecolor{SlateGray}{rgb}{0.44,0.50,0.56}
\definecolor{SlateGrey}{rgb}{0.44,0.50,0.56}
\definecolor{SpringGreen1}{rgb}{0.00,1.00,0.50}
\definecolor{SpringGreen2}{rgb}{0.00,0.93,0.46}
\definecolor{SpringGreen3}{rgb}{0.00,0.80,0.40}
\definecolor{SpringGreen4}{rgb}{0.00,0.55,0.27}
\definecolor{SpringGreen}{rgb}{0.00,1.00,0.50}
\definecolor{SteelBlue1}{rgb}{0.39,0.72,1.00}
\definecolor{SteelBlue2}{rgb}{0.36,0.67,0.93}
\definecolor{SteelBlue3}{rgb}{0.31,0.58,0.80}
\definecolor{SteelBlue4}{rgb}{0.21,0.39,0.55}
\definecolor{SteelBlue}{rgb}{0.27,0.51,0.71}
\definecolor{VioletRed1}{rgb}{1.00,0.24,0.59}
\definecolor{VioletRed2}{rgb}{0.93,0.23,0.55}
\definecolor{VioletRed3}{rgb}{0.80,0.20,0.47}
\definecolor{VioletRed4}{rgb}{0.55,0.13,0.32}
\definecolor{VioletRed}{rgb}{0.82,0.13,0.56}
\definecolor{WhiteSmoke}{rgb}{0.96,0.96,0.96}
\definecolor{YellowGreen}{rgb}{0.60,0.80,0.20}
\definecolor{aliceblue}{rgb}{0.94,0.97,1.00}
\definecolor{antiquewhite}{rgb}{0.98,0.92,0.84}
\definecolor{aquamarine1}{rgb}{0.50,1.00,0.83}
\definecolor{aquamarine2}{rgb}{0.46,0.93,0.78}
\definecolor{aquamarine3}{rgb}{0.40,0.80,0.67}
\definecolor{aquamarine4}{rgb}{0.27,0.55,0.45}
\definecolor{aquamarine}{rgb}{0.50,1.00,0.83}
\definecolor{azure1}{rgb}{0.94,1.00,1.00}
\definecolor{azure2}{rgb}{0.88,0.93,0.93}
\definecolor{azure3}{rgb}{0.76,0.80,0.80}
\definecolor{azure4}{rgb}{0.51,0.55,0.55}
\definecolor{azure}{rgb}{0.94,1.00,1.00}
\definecolor{beige}{rgb}{0.96,0.96,0.86}
\definecolor{bisque1}{rgb}{1.00,0.89,0.77}
\definecolor{bisque2}{rgb}{0.93,0.84,0.72}
\definecolor{bisque3}{rgb}{0.80,0.72,0.62}
\definecolor{bisque4}{rgb}{0.55,0.49,0.42}
\definecolor{bisque}{rgb}{1.00,0.89,0.77}
\definecolor{black}{rgb}{0.00,0.00,0.00}
\definecolor{blanchedalmond}{rgb}{1.00,0.92,0.80}
\definecolor{blue1}{rgb}{0.00,0.00,1.00}
\definecolor{blue2}{rgb}{0.00,0.00,0.93}
\definecolor{blue3}{rgb}{0.00,0.00,0.80}
\definecolor{blue4}{rgb}{0.00,0.00,0.55}
\definecolor{blueviolet}{rgb}{0.54,0.17,0.89}
\definecolor{blue}{rgb}{0.00,0.00,1.00}
\definecolor{brown1}{rgb}{1.00,0.25,0.25}
\definecolor{brown2}{rgb}{0.93,0.23,0.23}
\definecolor{brown3}{rgb}{0.80,0.20,0.20}
\definecolor{brown4}{rgb}{0.55,0.14,0.14}
\definecolor{brown}{rgb}{0.65,0.16,0.16}
\definecolor{burlywood1}{rgb}{1.00,0.83,0.61}
\definecolor{burlywood2}{rgb}{0.93,0.77,0.57}
\definecolor{burlywood3}{rgb}{0.80,0.67,0.49}
\definecolor{burlywood4}{rgb}{0.55,0.45,0.33}
\definecolor{burlywood}{rgb}{0.87,0.72,0.53}
\definecolor{cadetblue}{rgb}{0.37,0.62,0.63}
\definecolor{chartreuse1}{rgb}{0.50,1.00,0.00}
\definecolor{chartreuse2}{rgb}{0.46,0.93,0.00}
\definecolor{chartreuse3}{rgb}{0.40,0.80,0.00}
\definecolor{chartreuse4}{rgb}{0.27,0.55,0.00}
\definecolor{chartreuse}{rgb}{0.50,1.00,0.00}
\definecolor{chocolate1}{rgb}{1.00,0.50,0.14}
\definecolor{chocolate2}{rgb}{0.93,0.46,0.13}
\definecolor{chocolate3}{rgb}{0.80,0.40,0.11}
\definecolor{chocolate4}{rgb}{0.55,0.27,0.07}
\definecolor{chocolate}{rgb}{0.82,0.41,0.12}
\definecolor{coral1}{rgb}{1.00,0.45,0.34}
\definecolor{coral2}{rgb}{0.93,0.42,0.31}
\definecolor{coral3}{rgb}{0.80,0.36,0.27}
\definecolor{coral4}{rgb}{0.55,0.24,0.18}
\definecolor{coral}{rgb}{1.00,0.50,0.31}
\definecolor{cornflowerblue}{rgb}{0.39,0.58,0.93}
\definecolor{cornsilk1}{rgb}{1.00,0.97,0.86}
\definecolor{cornsilk2}{rgb}{0.93,0.91,0.80}
\definecolor{cornsilk3}{rgb}{0.80,0.78,0.69}
\definecolor{cornsilk4}{rgb}{0.55,0.53,0.47}
\definecolor{cornsilk}{rgb}{1.00,0.97,0.86}
\definecolor{cyan1}{rgb}{0.00,1.00,1.00}
\definecolor{cyan2}{rgb}{0.00,0.93,0.93}
\definecolor{cyan3}{rgb}{0.00,0.80,0.80}
\definecolor{cyan4}{rgb}{0.00,0.55,0.55}
\definecolor{cyan}{rgb}{0.00,1.00,1.00}
\definecolor{darkblue}{rgb}{0.00,0.00,0.55}
\definecolor{darkcyan}{rgb}{0.00,0.55,0.55}
\definecolor{darkgoldenrod}{rgb}{0.72,0.53,0.04}
\definecolor{darkgray}{rgb}{0.66,0.66,0.66}
\definecolor{darkgreen}{rgb}{0.00,0.39,0.00}
\definecolor{darkgrey}{rgb}{0.66,0.66,0.66}
\definecolor{darkkhaki}{rgb}{0.74,0.72,0.42}
\definecolor{darkmagenta}{rgb}{0.55,0.00,0.55}
\definecolor{darkolive}{rgb}{0.33,0.42,0.18}
\definecolor{darkorange}{rgb}{1.00,0.55,0.00}
\definecolor{darkorchid}{rgb}{0.60,0.20,0.80}
\definecolor{darkred}{rgb}{0.55,0.00,0.00}
\definecolor{darksalmon}{rgb}{0.91,0.59,0.48}
\definecolor{darksea}{rgb}{0.56,0.74,0.56}
\definecolor{darkslate}{rgb}{0.18,0.31,0.31}
\definecolor{darkslate}{rgb}{0.18,0.31,0.31}
\definecolor{darkslate}{rgb}{0.28,0.24,0.55}
\definecolor{darkturquoise}{rgb}{0.00,0.81,0.82}
\definecolor{darkviolet}{rgb}{0.58,0.00,0.83}
\definecolor{deeppink}{rgb}{1.00,0.08,0.58}
\definecolor{deepsky}{rgb}{0.00,0.75,1.00}
\definecolor{dimgray}{rgb}{0.41,0.41,0.41}
\definecolor{dimgrey}{rgb}{0.41,0.41,0.41}
\definecolor{dodgerblue}{rgb}{0.12,0.56,1.00}
\definecolor{firebrick1}{rgb}{1.00,0.19,0.19}
\definecolor{firebrick2}{rgb}{0.93,0.17,0.17}
\definecolor{firebrick3}{rgb}{0.80,0.15,0.15}
\definecolor{firebrick4}{rgb}{0.55,0.10,0.10}
\definecolor{firebrick}{rgb}{0.70,0.13,0.13}
\definecolor{floralwhite}{rgb}{1.00,0.98,0.94}
\definecolor{forestgreen}{rgb}{0.13,0.55,0.13}
\definecolor{gainsboro}{rgb}{0.86,0.86,0.86}
\definecolor{ghostwhite}{rgb}{0.97,0.97,1.00}
\definecolor{gold1}{rgb}{1.00,0.84,0.00}
\definecolor{gold2}{rgb}{0.93,0.79,0.00}
\definecolor{gold3}{rgb}{0.80,0.68,0.00}
\definecolor{gold4}{rgb}{0.55,0.46,0.00}
\definecolor{goldenrod1}{rgb}{1.00,0.76,0.15}
\definecolor{goldenrod2}{rgb}{0.93,0.71,0.13}
\definecolor{goldenrod3}{rgb}{0.80,0.61,0.11}
\definecolor{goldenrod4}{rgb}{0.55,0.41,0.08}
\definecolor{goldenrod}{rgb}{0.85,0.65,0.13}
\definecolor{gold}{rgb}{1.00,0.84,0.00}
\definecolor{gray0}{rgb}{0.00,0.00,0.00}
\definecolor{gray100}{rgb}{1.00,1.00,1.00}
\definecolor{gray10}{rgb}{0.10,0.10,0.10}
\definecolor{gray11}{rgb}{0.11,0.11,0.11}
\definecolor{gray12}{rgb}{0.12,0.12,0.12}
\definecolor{gray13}{rgb}{0.13,0.13,0.13}
\definecolor{gray14}{rgb}{0.14,0.14,0.14}
\definecolor{gray15}{rgb}{0.15,0.15,0.15}
\definecolor{gray16}{rgb}{0.16,0.16,0.16}
\definecolor{gray17}{rgb}{0.17,0.17,0.17}
\definecolor{gray18}{rgb}{0.18,0.18,0.18}
\definecolor{gray19}{rgb}{0.19,0.19,0.19}
\definecolor{gray1}{rgb}{0.01,0.01,0.01}
\definecolor{gray20}{rgb}{0.20,0.20,0.20}
\definecolor{gray21}{rgb}{0.21,0.21,0.21}
\definecolor{gray22}{rgb}{0.22,0.22,0.22}
\definecolor{gray23}{rgb}{0.23,0.23,0.23}
\definecolor{gray24}{rgb}{0.24,0.24,0.24}
\definecolor{gray25}{rgb}{0.25,0.25,0.25}
\definecolor{gray26}{rgb}{0.26,0.26,0.26}
\definecolor{gray27}{rgb}{0.27,0.27,0.27}
\definecolor{gray28}{rgb}{0.28,0.28,0.28}
\definecolor{gray29}{rgb}{0.29,0.29,0.29}
\definecolor{gray2}{rgb}{0.02,0.02,0.02}
\definecolor{gray30}{rgb}{0.30,0.30,0.30}
\definecolor{gray31}{rgb}{0.31,0.31,0.31}
\definecolor{gray32}{rgb}{0.32,0.32,0.32}
\definecolor{gray33}{rgb}{0.33,0.33,0.33}
\definecolor{gray34}{rgb}{0.34,0.34,0.34}
\definecolor{gray35}{rgb}{0.35,0.35,0.35}
\definecolor{gray36}{rgb}{0.36,0.36,0.36}
\definecolor{gray37}{rgb}{0.37,0.37,0.37}
\definecolor{gray38}{rgb}{0.38,0.38,0.38}
\definecolor{gray39}{rgb}{0.39,0.39,0.39}
\definecolor{gray3}{rgb}{0.03,0.03,0.03}
\definecolor{gray40}{rgb}{0.40,0.40,0.40}
\definecolor{gray41}{rgb}{0.41,0.41,0.41}
\definecolor{gray42}{rgb}{0.42,0.42,0.42}
\definecolor{gray43}{rgb}{0.43,0.43,0.43}
\definecolor{gray44}{rgb}{0.44,0.44,0.44}
\definecolor{gray45}{rgb}{0.45,0.45,0.45}
\definecolor{gray46}{rgb}{0.46,0.46,0.46}
\definecolor{gray47}{rgb}{0.47,0.47,0.47}
\definecolor{gray48}{rgb}{0.48,0.48,0.48}
\definecolor{gray49}{rgb}{0.49,0.49,0.49}
\definecolor{gray4}{rgb}{0.04,0.04,0.04}
\definecolor{gray50}{rgb}{0.50,0.50,0.50}
\definecolor{gray51}{rgb}{0.51,0.51,0.51}
\definecolor{gray52}{rgb}{0.52,0.52,0.52}
\definecolor{gray53}{rgb}{0.53,0.53,0.53}
\definecolor{gray54}{rgb}{0.54,0.54,0.54}
\definecolor{gray55}{rgb}{0.55,0.55,0.55}
\definecolor{gray56}{rgb}{0.56,0.56,0.56}
\definecolor{gray57}{rgb}{0.57,0.57,0.57}
\definecolor{gray58}{rgb}{0.58,0.58,0.58}
\definecolor{gray59}{rgb}{0.59,0.59,0.59}
\definecolor{gray5}{rgb}{0.05,0.05,0.05}
\definecolor{gray60}{rgb}{0.60,0.60,0.60}
\definecolor{gray61}{rgb}{0.61,0.61,0.61}
\definecolor{gray62}{rgb}{0.62,0.62,0.62}
\definecolor{gray63}{rgb}{0.63,0.63,0.63}
\definecolor{gray64}{rgb}{0.64,0.64,0.64}
\definecolor{gray65}{rgb}{0.65,0.65,0.65}
\definecolor{gray66}{rgb}{0.66,0.66,0.66}
\definecolor{gray67}{rgb}{0.67,0.67,0.67}
\definecolor{gray68}{rgb}{0.68,0.68,0.68}
\definecolor{gray69}{rgb}{0.69,0.69,0.69}
\definecolor{gray6}{rgb}{0.06,0.06,0.06}
\definecolor{gray70}{rgb}{0.70,0.70,0.70}
\definecolor{gray71}{rgb}{0.71,0.71,0.71}
\definecolor{gray72}{rgb}{0.72,0.72,0.72}
\definecolor{gray73}{rgb}{0.73,0.73,0.73}
\definecolor{gray74}{rgb}{0.74,0.74,0.74}
\definecolor{gray75}{rgb}{0.75,0.75,0.75}
\definecolor{gray76}{rgb}{0.76,0.76,0.76}
\definecolor{gray77}{rgb}{0.77,0.77,0.77}
\definecolor{gray78}{rgb}{0.78,0.78,0.78}
\definecolor{gray79}{rgb}{0.79,0.79,0.79}
\definecolor{gray7}{rgb}{0.07,0.07,0.07}
\definecolor{gray80}{rgb}{0.80,0.80,0.80}
\definecolor{gray81}{rgb}{0.81,0.81,0.81}
\definecolor{gray82}{rgb}{0.82,0.82,0.82}
\definecolor{gray83}{rgb}{0.83,0.83,0.83}
\definecolor{gray84}{rgb}{0.84,0.84,0.84}
\definecolor{gray85}{rgb}{0.85,0.85,0.85}
\definecolor{gray86}{rgb}{0.86,0.86,0.86}
\definecolor{gray87}{rgb}{0.87,0.87,0.87}
\definecolor{gray88}{rgb}{0.88,0.88,0.88}
\definecolor{gray89}{rgb}{0.89,0.89,0.89}
\definecolor{gray8}{rgb}{0.08,0.08,0.08}
\definecolor{gray90}{rgb}{0.90,0.90,0.90}
\definecolor{gray91}{rgb}{0.91,0.91,0.91}
\definecolor{gray92}{rgb}{0.92,0.92,0.92}
\definecolor{gray93}{rgb}{0.93,0.93,0.93}
\definecolor{gray94}{rgb}{0.94,0.94,0.94}
\definecolor{gray95}{rgb}{0.95,0.95,0.95}
\definecolor{gray96}{rgb}{0.96,0.96,0.96}
\definecolor{gray97}{rgb}{0.97,0.97,0.97}
\definecolor{gray98}{rgb}{0.98,0.98,0.98}
\definecolor{gray99}{rgb}{0.99,0.99,0.99}
\definecolor{gray9}{rgb}{0.09,0.09,0.09}
\definecolor{gray}{rgb}{0.75,0.75,0.75}
\definecolor{green1}{rgb}{0.00,1.00,0.00}
\definecolor{green2}{rgb}{0.00,0.93,0.00}
\definecolor{green3}{rgb}{0.00,0.80,0.00}
\definecolor{green4}{rgb}{0.00,0.55,0.00}
\definecolor{greenyellow}{rgb}{0.68,1.00,0.18}
\definecolor{green}{rgb}{0.00,1.00,0.00}
\definecolor{grey0}{rgb}{0.00,0.00,0.00}
\definecolor{grey100}{rgb}{1.00,1.00,1.00}
\definecolor{grey10}{rgb}{0.10,0.10,0.10}
\definecolor{grey11}{rgb}{0.11,0.11,0.11}
\definecolor{grey12}{rgb}{0.12,0.12,0.12}
\definecolor{grey13}{rgb}{0.13,0.13,0.13}
\definecolor{grey14}{rgb}{0.14,0.14,0.14}
\definecolor{grey15}{rgb}{0.15,0.15,0.15}
\definecolor{grey16}{rgb}{0.16,0.16,0.16}
\definecolor{grey17}{rgb}{0.17,0.17,0.17}
\definecolor{grey18}{rgb}{0.18,0.18,0.18}
\definecolor{grey19}{rgb}{0.19,0.19,0.19}
\definecolor{grey1}{rgb}{0.01,0.01,0.01}
\definecolor{grey20}{rgb}{0.20,0.20,0.20}
\definecolor{grey21}{rgb}{0.21,0.21,0.21}
\definecolor{grey22}{rgb}{0.22,0.22,0.22}
\definecolor{grey23}{rgb}{0.23,0.23,0.23}
\definecolor{grey24}{rgb}{0.24,0.24,0.24}
\definecolor{grey25}{rgb}{0.25,0.25,0.25}
\definecolor{grey26}{rgb}{0.26,0.26,0.26}
\definecolor{grey27}{rgb}{0.27,0.27,0.27}
\definecolor{grey28}{rgb}{0.28,0.28,0.28}
\definecolor{grey29}{rgb}{0.29,0.29,0.29}
\definecolor{grey2}{rgb}{0.02,0.02,0.02}
\definecolor{grey30}{rgb}{0.30,0.30,0.30}
\definecolor{grey31}{rgb}{0.31,0.31,0.31}
\definecolor{grey32}{rgb}{0.32,0.32,0.32}
\definecolor{grey33}{rgb}{0.33,0.33,0.33}
\definecolor{grey34}{rgb}{0.34,0.34,0.34}
\definecolor{grey35}{rgb}{0.35,0.35,0.35}
\definecolor{grey36}{rgb}{0.36,0.36,0.36}
\definecolor{grey37}{rgb}{0.37,0.37,0.37}
\definecolor{grey38}{rgb}{0.38,0.38,0.38}
\definecolor{grey39}{rgb}{0.39,0.39,0.39}
\definecolor{grey3}{rgb}{0.03,0.03,0.03}
\definecolor{grey40}{rgb}{0.40,0.40,0.40}
\definecolor{grey41}{rgb}{0.41,0.41,0.41}
\definecolor{grey42}{rgb}{0.42,0.42,0.42}
\definecolor{grey43}{rgb}{0.43,0.43,0.43}
\definecolor{grey44}{rgb}{0.44,0.44,0.44}
\definecolor{grey45}{rgb}{0.45,0.45,0.45}
\definecolor{grey46}{rgb}{0.46,0.46,0.46}
\definecolor{grey47}{rgb}{0.47,0.47,0.47}
\definecolor{grey48}{rgb}{0.48,0.48,0.48}
\definecolor{grey49}{rgb}{0.49,0.49,0.49}
\definecolor{grey4}{rgb}{0.04,0.04,0.04}
\definecolor{grey50}{rgb}{0.50,0.50,0.50}
\definecolor{grey51}{rgb}{0.51,0.51,0.51}
\definecolor{grey52}{rgb}{0.52,0.52,0.52}
\definecolor{grey53}{rgb}{0.53,0.53,0.53}
\definecolor{grey54}{rgb}{0.54,0.54,0.54}
\definecolor{grey55}{rgb}{0.55,0.55,0.55}
\definecolor{grey56}{rgb}{0.56,0.56,0.56}
\definecolor{grey57}{rgb}{0.57,0.57,0.57}
\definecolor{grey58}{rgb}{0.58,0.58,0.58}
\definecolor{grey59}{rgb}{0.59,0.59,0.59}
\definecolor{grey5}{rgb}{0.05,0.05,0.05}
\definecolor{grey60}{rgb}{0.60,0.60,0.60}
\definecolor{grey61}{rgb}{0.61,0.61,0.61}
\definecolor{grey62}{rgb}{0.62,0.62,0.62}
\definecolor{grey63}{rgb}{0.63,0.63,0.63}
\definecolor{grey64}{rgb}{0.64,0.64,0.64}
\definecolor{grey65}{rgb}{0.65,0.65,0.65}
\definecolor{grey66}{rgb}{0.66,0.66,0.66}
\definecolor{grey67}{rgb}{0.67,0.67,0.67}
\definecolor{grey68}{rgb}{0.68,0.68,0.68}
\definecolor{grey69}{rgb}{0.69,0.69,0.69}
\definecolor{grey6}{rgb}{0.06,0.06,0.06}
\definecolor{grey70}{rgb}{0.70,0.70,0.70}
\definecolor{grey71}{rgb}{0.71,0.71,0.71}
\definecolor{grey72}{rgb}{0.72,0.72,0.72}
\definecolor{grey73}{rgb}{0.73,0.73,0.73}
\definecolor{grey74}{rgb}{0.74,0.74,0.74}
\definecolor{grey75}{rgb}{0.75,0.75,0.75}
\definecolor{grey76}{rgb}{0.76,0.76,0.76}
\definecolor{grey77}{rgb}{0.77,0.77,0.77}
\definecolor{grey78}{rgb}{0.78,0.78,0.78}
\definecolor{grey79}{rgb}{0.79,0.79,0.79}
\definecolor{grey7}{rgb}{0.07,0.07,0.07}
\definecolor{grey80}{rgb}{0.80,0.80,0.80}
\definecolor{grey81}{rgb}{0.81,0.81,0.81}
\definecolor{grey82}{rgb}{0.82,0.82,0.82}
\definecolor{grey83}{rgb}{0.83,0.83,0.83}
\definecolor{grey84}{rgb}{0.84,0.84,0.84}
\definecolor{grey85}{rgb}{0.85,0.85,0.85}
\definecolor{grey86}{rgb}{0.86,0.86,0.86}
\definecolor{grey87}{rgb}{0.87,0.87,0.87}
\definecolor{grey88}{rgb}{0.88,0.88,0.88}
\definecolor{grey89}{rgb}{0.89,0.89,0.89}
\definecolor{grey8}{rgb}{0.08,0.08,0.08}
\definecolor{grey90}{rgb}{0.90,0.90,0.90}
\definecolor{grey91}{rgb}{0.91,0.91,0.91}
\definecolor{grey92}{rgb}{0.92,0.92,0.92}
\definecolor{grey93}{rgb}{0.93,0.93,0.93}
\definecolor{grey94}{rgb}{0.94,0.94,0.94}
\definecolor{grey95}{rgb}{0.95,0.95,0.95}
\definecolor{grey96}{rgb}{0.96,0.96,0.96}
\definecolor{grey97}{rgb}{0.97,0.97,0.97}
\definecolor{grey98}{rgb}{0.98,0.98,0.98}
\definecolor{grey99}{rgb}{0.99,0.99,0.99}
\definecolor{grey9}{rgb}{0.09,0.09,0.09}
\definecolor{grey}{rgb}{0.75,0.75,0.75}
\definecolor{honeydew1}{rgb}{0.94,1.00,0.94}
\definecolor{honeydew2}{rgb}{0.88,0.93,0.88}
\definecolor{honeydew3}{rgb}{0.76,0.80,0.76}
\definecolor{honeydew4}{rgb}{0.51,0.55,0.51}
\definecolor{honeydew}{rgb}{0.94,1.00,0.94}
\definecolor{hotpink}{rgb}{1.00,0.41,0.71}
\definecolor{indianred}{rgb}{0.80,0.36,0.36}
\definecolor{ivory1}{rgb}{1.00,1.00,0.94}
\definecolor{ivory2}{rgb}{0.93,0.93,0.88}
\definecolor{ivory3}{rgb}{0.80,0.80,0.76}
\definecolor{ivory4}{rgb}{0.55,0.55,0.51}
\definecolor{ivory}{rgb}{1.00,1.00,0.94}
\definecolor{khaki1}{rgb}{1.00,0.96,0.56}
\definecolor{khaki2}{rgb}{0.93,0.90,0.52}
\definecolor{khaki3}{rgb}{0.80,0.78,0.45}
\definecolor{khaki4}{rgb}{0.55,0.53,0.31}
\definecolor{khaki}{rgb}{0.94,0.90,0.55}
\definecolor{lavenderblush}{rgb}{1.00,0.94,0.96}
\definecolor{lavender}{rgb}{0.90,0.90,0.98}
\definecolor{lawngreen}{rgb}{0.49,0.99,0.00}
\definecolor{lemonchiffon}{rgb}{1.00,0.98,0.80}
\definecolor{lightblue}{rgb}{0.68,0.85,0.90}
\definecolor{lightcoral}{rgb}{0.94,0.50,0.50}
\definecolor{lightcyan}{rgb}{0.88,1.00,1.00}
\definecolor{lightgoldenrod}{rgb}{0.93,0.87,0.51}
\definecolor{lightgoldenrod}{rgb}{0.98,0.98,0.82}
\definecolor{lightgray}{rgb}{0.83,0.83,0.83}
\definecolor{lightgreen}{rgb}{0.56,0.93,0.56}
\definecolor{lightgrey}{rgb}{0.83,0.83,0.83}
\definecolor{lightpink}{rgb}{1.00,0.71,0.76}
\definecolor{lightsalmon}{rgb}{1.00,0.63,0.48}
\definecolor{lightsea}{rgb}{0.13,0.70,0.67}
\definecolor{lightsky}{rgb}{0.53,0.81,0.98}
\definecolor{lightslate}{rgb}{0.47,0.53,0.60}
\definecolor{lightslate}{rgb}{0.47,0.53,0.60}
\definecolor{lightslate}{rgb}{0.52,0.44,1.00}
\definecolor{lightsteel}{rgb}{0.69,0.77,0.87}
\definecolor{lightyellow}{rgb}{1.00,1.00,0.88}
\definecolor{limegreen}{rgb}{0.20,0.80,0.20}
\definecolor{linen}{rgb}{0.98,0.94,0.90}
\definecolor{magenta1}{rgb}{1.00,0.00,1.00}
\definecolor{magenta2}{rgb}{0.93,0.00,0.93}
\definecolor{magenta3}{rgb}{0.80,0.00,0.80}
\definecolor{magenta4}{rgb}{0.55,0.00,0.55}
\definecolor{magenta}{rgb}{1.00,0.00,1.00}
\definecolor{maroon1}{rgb}{1.00,0.20,0.70}
\definecolor{maroon2}{rgb}{0.93,0.19,0.65}
\definecolor{maroon3}{rgb}{0.80,0.16,0.56}
\definecolor{maroon4}{rgb}{0.55,0.11,0.38}
\definecolor{maroon}{rgb}{0.69,0.19,0.38}
\definecolor{mediumaquamarine}{rgb}{0.40,0.80,0.67}
\definecolor{mediumblue}{rgb}{0.00,0.00,0.80}
\definecolor{mediumorchid}{rgb}{0.73,0.33,0.83}
\definecolor{mediumpurple}{rgb}{0.58,0.44,0.86}
\definecolor{mediumsea}{rgb}{0.24,0.70,0.44}
\definecolor{mediumslate}{rgb}{0.48,0.41,0.93}
\definecolor{mediumspring}{rgb}{0.00,0.98,0.60}
\definecolor{mediumturquoise}{rgb}{0.28,0.82,0.80}
\definecolor{mediumviolet}{rgb}{0.78,0.08,0.52}
\definecolor{midnightblue}{rgb}{0.10,0.10,0.44}
\definecolor{mintcream}{rgb}{0.96,1.00,0.98}
\definecolor{mistyrose}{rgb}{1.00,0.89,0.88}
\definecolor{moccasin}{rgb}{1.00,0.89,0.71}
\definecolor{navajowhite}{rgb}{1.00,0.87,0.68}
\definecolor{navyblue}{rgb}{0.00,0.00,0.50}
\definecolor{navy}{rgb}{0.00,0.00,0.50}
\definecolor{oldlace}{rgb}{0.99,0.96,0.90}
\definecolor{olivedrab}{rgb}{0.42,0.56,0.14}
\definecolor{orange1}{rgb}{1.00,0.65,0.00}
\definecolor{orange2}{rgb}{0.93,0.60,0.00}
\definecolor{orange3}{rgb}{0.80,0.52,0.00}
\definecolor{orange4}{rgb}{0.55,0.35,0.00}
\definecolor{orangered}{rgb}{1.00,0.27,0.00}
\definecolor{orange}{rgb}{1.00,0.65,0.00}
\definecolor{orchid1}{rgb}{1.00,0.51,0.98}
\definecolor{orchid2}{rgb}{0.93,0.48,0.91}
\definecolor{orchid3}{rgb}{0.80,0.41,0.79}
\definecolor{orchid4}{rgb}{0.55,0.28,0.54}
\definecolor{orchid}{rgb}{0.85,0.44,0.84}
\definecolor{palegoldenrod}{rgb}{0.93,0.91,0.67}
\definecolor{palegreen}{rgb}{0.60,0.98,0.60}
\definecolor{paleturquoise}{rgb}{0.69,0.93,0.93}
\definecolor{paleviolet}{rgb}{0.86,0.44,0.58}
\definecolor{papayawhip}{rgb}{1.00,0.94,0.84}
\definecolor{peachpuff}{rgb}{1.00,0.85,0.73}
\definecolor{peru}{rgb}{0.80,0.52,0.25}
\definecolor{pink1}{rgb}{1.00,0.71,0.77}
\definecolor{pink2}{rgb}{0.93,0.66,0.72}
\definecolor{pink3}{rgb}{0.80,0.57,0.62}
\definecolor{pink4}{rgb}{0.55,0.39,0.42}
\definecolor{pink}{rgb}{1.00,0.75,0.80}
\definecolor{plum1}{rgb}{1.00,0.73,1.00}
\definecolor{plum2}{rgb}{0.93,0.68,0.93}
\definecolor{plum3}{rgb}{0.80,0.59,0.80}
\definecolor{plum4}{rgb}{0.55,0.40,0.55}
\definecolor{plum}{rgb}{0.87,0.63,0.87}
\definecolor{powderblue}{rgb}{0.69,0.88,0.90}
\definecolor{purple1}{rgb}{0.61,0.19,1.00}
\definecolor{purple2}{rgb}{0.57,0.17,0.93}
\definecolor{purple3}{rgb}{0.49,0.15,0.80}
\definecolor{purple4}{rgb}{0.33,0.10,0.55}
\definecolor{purple}{rgb}{0.63,0.13,0.94}
\definecolor{red1}{rgb}{1.00,0.00,0.00}
\definecolor{red2}{rgb}{0.93,0.00,0.00}
\definecolor{red3}{rgb}{0.80,0.00,0.00}
\definecolor{red4}{rgb}{0.55,0.00,0.00}
\definecolor{red}{rgb}{1.00,0.00,0.00}
\definecolor{rosybrown}{rgb}{0.74,0.56,0.56}
\definecolor{royalblue}{rgb}{0.25,0.41,0.88}
\definecolor{saddlebrown}{rgb}{0.55,0.27,0.07}
\definecolor{salmon1}{rgb}{1.00,0.55,0.41}
\definecolor{salmon2}{rgb}{0.93,0.51,0.38}
\definecolor{salmon3}{rgb}{0.80,0.44,0.33}
\definecolor{salmon4}{rgb}{0.55,0.30,0.22}
\definecolor{salmon}{rgb}{0.98,0.50,0.45}
\definecolor{sandybrown}{rgb}{0.96,0.64,0.38}
\definecolor{seagreen}{rgb}{0.18,0.55,0.34}
\definecolor{seashell1}{rgb}{1.00,0.96,0.93}
\definecolor{seashell2}{rgb}{0.93,0.90,0.87}
\definecolor{seashell3}{rgb}{0.80,0.77,0.75}
\definecolor{seashell4}{rgb}{0.55,0.53,0.51}
\definecolor{seashell}{rgb}{1.00,0.96,0.93}
\definecolor{sienna1}{rgb}{1.00,0.51,0.28}
\definecolor{sienna2}{rgb}{0.93,0.47,0.26}
\definecolor{sienna3}{rgb}{0.80,0.41,0.22}
\definecolor{sienna4}{rgb}{0.55,0.28,0.15}
\definecolor{sienna}{rgb}{0.63,0.32,0.18}
\definecolor{skyblue}{rgb}{0.53,0.81,0.92}
\definecolor{slateblue}{rgb}{0.42,0.35,0.80}
\definecolor{slategray}{rgb}{0.44,0.50,0.56}
\definecolor{slategrey}{rgb}{0.44,0.50,0.56}
\definecolor{snow1}{rgb}{1.00,0.98,0.98}
\definecolor{snow2}{rgb}{0.93,0.91,0.91}
\definecolor{snow3}{rgb}{0.80,0.79,0.79}
\definecolor{snow4}{rgb}{0.55,0.54,0.54}
\definecolor{snow}{rgb}{1.00,0.98,0.98}
\definecolor{springgreen}{rgb}{0.00,1.00,0.50}
\definecolor{steelblue}{rgb}{0.27,0.51,0.71}
\definecolor{tan1}{rgb}{1.00,0.65,0.31}
\definecolor{tan2}{rgb}{0.93,0.60,0.29}
\definecolor{tan3}{rgb}{0.80,0.52,0.25}
\definecolor{tan4}{rgb}{0.55,0.35,0.17}
\definecolor{tan}{rgb}{0.82,0.71,0.55}
\definecolor{thistle1}{rgb}{1.00,0.88,1.00}
\definecolor{thistle2}{rgb}{0.93,0.82,0.93}
\definecolor{thistle3}{rgb}{0.80,0.71,0.80}
\definecolor{thistle4}{rgb}{0.55,0.48,0.55}
\definecolor{thistle}{rgb}{0.85,0.75,0.85}
\definecolor{tomato1}{rgb}{1.00,0.39,0.28}
\definecolor{tomato2}{rgb}{0.93,0.36,0.26}
\definecolor{tomato3}{rgb}{0.80,0.31,0.22}
\definecolor{tomato4}{rgb}{0.55,0.21,0.15}
\definecolor{tomato}{rgb}{1.00,0.39,0.28}
\definecolor{turquoise1}{rgb}{0.00,0.96,1.00}
\definecolor{turquoise2}{rgb}{0.00,0.90,0.93}
\definecolor{turquoise3}{rgb}{0.00,0.77,0.80}
\definecolor{turquoise4}{rgb}{0.00,0.53,0.55}
\definecolor{turquoise}{rgb}{0.25,0.88,0.82}
\definecolor{violetred}{rgb}{0.82,0.13,0.56}
\definecolor{violet}{rgb}{0.93,0.51,0.93}
\definecolor{wheat1}{rgb}{1.00,0.91,0.73}
\definecolor{wheat2}{rgb}{0.93,0.85,0.68}
\definecolor{wheat3}{rgb}{0.80,0.73,0.59}
\definecolor{wheat4}{rgb}{0.55,0.49,0.40}
\definecolor{wheat}{rgb}{0.96,0.87,0.70}
\definecolor{whitesmoke}{rgb}{0.96,0.96,0.96}
\definecolor{white}{rgb}{1.00,1.00,1.00}
\definecolor{yellow1}{rgb}{1.00,1.00,0.00}
\definecolor{yellow2}{rgb}{0.93,0.93,0.00}
\definecolor{yellow3}{rgb}{0.80,0.80,0.00}
\definecolor{yellow4}{rgb}{0.55,0.55,0.00}
\definecolor{yellowgreen}{rgb}{0.60,0.80,0.20}
\definecolor{yellow}{rgb}{1.00,1.00,0.00}

%% file: cbi_instr.bbl
\begin{thebibliography}{53}
\expandafter\ifx\csname natexlab\endcsname\relax\def\natexlab#1{#1}\fi

\bibitem[{{Benson} {et~al}\mbox{.}(2004){Benson}, {Church}, {Ade}, {Bock},
  {Ganga}, {Henson}, \& {Thompson}}]{Benson:2004}
{Benson} B.~A., {Church} S.~E., {Ade} P.~A.~R., {Bock} J.~J., {Ganga} K.~M.,
  {Henson} C.~N., {Thompson} K.~L., 2004, ApJ, 617, 829

\bibitem[{{Benson} {et~al}\mbox{.}(2003){Benson}, {Church}, {Ade}, {Bock},
  {Ganga}, {Hinderks}, {Mauskopf}, {Philhour}, {Runyan}, \&
  {Thompson}}]{Benson:2003}
{Benson} B.~A. {et~al.}, 2003, ApJ, 592, 674

\bibitem[{{Bonamente} {et~al}\mbox{.}(2008){Bonamente}, {Joy}, {LaRoque},
  {Carlstrom}, {Nagai}, \& {Marrone}}]{Bonamente:2008}
{Bonamente} M., {Joy} M., {LaRoque} S.~J., {Carlstrom} J.~E., {Nagai} D.,
  {Marrone} D.~P., 2008, ApJ, 675, 106

\bibitem[{{Casassus} {et~al}\mbox{.}(2006){Casassus}, {Cabrera}, {F{\"o}rster},
  {Pearson}, {Readhead}, \& {Dickinson}}]{Casassus:2006}
{Casassus} S., {Cabrera} G.~F., {F{\"o}rster} F., {Pearson} T.~J., {Readhead}
  A.~C.~S., {Dickinson} C., 2006, ApJ, 639, 951

\bibitem[{{Casassus} {et~al}\mbox{.}(2008){Casassus}, {Dickinson}, {Cleary},
  {Paladini}, {Etxaluze}, {Lim}, {White}, {Burton}, {Indermuehle}, {Stahl}, \&
  {Roche}}]{Casassus:2008}
{Casassus} S. {et~al.}, 2008, MNRAS, 391, 1075

\bibitem[{{Casassus} {et~al}\mbox{.}(2004){Casassus}, {Readhead}, {Pearson},
  {Nyman}, {Shepherd}, \& {Bronfman}}]{Casassus:2004}
{Casassus} S., {Readhead} A.~C.~S., {Pearson} T.~J., {Nyman} L., {Shepherd}
  M.~C., {Bronfman} L., 2004, ApJ, 603, 599

\bibitem[{{Castellanos} {et~al}\mbox{.}(2011){Castellanos}, {Casassus},
  {Dickinson}, {Vidal}, {Paladini}, {Cleary}, {Davies}, {Davis}, {White}, \&
  {Taylor}}]{Castellanos:2010}
{Castellanos} P. {et~al.}, 2011, MNRAS, 411, 1137

\bibitem[{{Cavagnolo} {et~al}\mbox{.}(2009){Cavagnolo}, {Donahue}, {Voit}, \&
  {Sun}}]{Cavagnolo:2009}
{Cavagnolo} K.~W., {Donahue} M., {Voit} G.~M., {Sun} M., 2009, ApJS, 182, 12

\bibitem[{{Cavaliere} \& {Fusco-Femiano}(1976)}]{Cavaliere:1976}
{Cavaliere} A., {Fusco-Femiano} R., 1976, A\&A, 49, 137

\bibitem[{{Cavaliere} \& {Fusco-Femiano}(1978)}]{Cavaliere:1978}
---, 1978, A\&A, 70, 677

\bibitem[{{Colberg} {et~al}\mbox{.}(2000){Colberg}, {White}, {MacFarland},
  {Jenkins}, {Pearce}, {Frenk}, {Thomas}, \& {Couchman}}]{Colberg:2000}
{Colberg} J.~M., {White} S.~D.~M., {MacFarland} T.~J., {Jenkins} A., {Pearce}
  F.~R., {Frenk} C.~S., {Thomas} P.~A., {Couchman} H.~M.~P., 2000, MNRAS, 313,
  229

\bibitem[{{Dale} {et~al}\mbox{.}(1999){Dale}, {Giovanelli}, {Haynes},
  {Campusano}, \& {Hardy}}]{Dale:1999}
{Dale} D.~A., {Giovanelli} R., {Haynes} M.~P., {Campusano} L.~E., {Hardy} E.,
  1999, AJ, 118, 1489

\bibitem[{{Dickinson} {et~al}\mbox{.}(2010){Dickinson}, {Casassus}, {Davies},
  {Allison}, {Bustos}, {Cleary}, {Davis}, {Jones}, {Pearson}, {Readhead},
  {Reeves}, {Taylor}, {Tibbs}, \& {Watson}}]{Dickinson:2010}
{Dickinson} C. {et~al.}, 2010, MNRAS, 407, 2223

\bibitem[{{Dickinson} {et~al}\mbox{.}(2006){Dickinson}, {Casassus}, {Pineda},
  {Pearson}, {Readhead}, \& {Davies}}]{Dickinson:2006}
{Dickinson} C., {Casassus} S., {Pineda} J.~L., {Pearson} T.~J., {Readhead}
  A.~C.~S., {Davies} R.~D., 2006, ApJL, 643, L111

\bibitem[{{Dickinson} {et~al}\mbox{.}(2009){Dickinson}, {Davies}, {Allison},
  {Bond}, {Casassus}, {Cleary}, {Davis}, {Jones}, {Mason}, {Myers}, {Pearson},
  {Readhead}, {Sievers}, {Taylor}, {Todorovi{\'c}}, {White}, \&
  {Wilkinson}}]{Dickinson:2009}
{Dickinson} C. {et~al.}, 2009, ApJ, 690, 1585

\bibitem[{{Dickinson} {et~al}\mbox{.}(2007){Dickinson}, {Davies}, {Bronfman},
  {Casassus}, {Davis}, {Pearson}, {Readhead}, \& {Wilkinson}}]{Dickinson:2007}
{Dickinson} C., {Davies} R.~D., {Bronfman} L., {Casassus} S., {Davis} R.~J.,
  {Pearson} T.~J., {Readhead} A.~C.~S., {Wilkinson} P.~N., 2007, MNRAS, 379,
  297

\bibitem[{{Feroz}, {Hobson} \& {Bridges}(2009){Feroz}, {Hobson}, \&
  {Bridges}}]{Feroz:2009}
{Feroz} F., {Hobson} M.~P., {Bridges} M., 2009, MNRAS, 398, 1601

\bibitem[{{Giovanelli} {et~al}\mbox{.}(1998){Giovanelli}, {Haynes}, {Salzer},
  {Wegner}, {da Costa}, \& {Freudling}}]{Giovanelli:1998}
{Giovanelli} R., {Haynes} M.~P., {Salzer} J.~J., {Wegner} G., {da Costa} L.~N.,
  {Freudling} W., 1998, AJ, 116, 2632

\bibitem[{{Hales} {et~al}\mbox{.}(2004){Hales}, {Casassus}, {Alvarez}, {May},
  {Bronfman}, {Readhead}, {Pearson}, {Mason}, \& {Dodson}}]{Hales:2004}
{Hales} A.~S. {et~al.}, 2004, ApJ, 613, 977

\bibitem[{{Hill} {et~al}\mbox{.}(2009){Hill}, {Weiland}, {Odegard}, {Wollack},
  {Hinshaw}, {Larson}, {Bennett}, {Halpern}, {Page}, {Dunkley}, {Gold},
  {Jarosik}, {Kogut}, {Limon}, {Nolta}, {Spergel}, {Tucker}, \&
  {Wright}}]{Hill:2009}
{Hill} R.~S. {et~al.}, 2009, ApJS, 180, 246

\bibitem[{{Ho} {et~al}\mbox{.}(2009){Ho}, {Altamirano}, {Chang}, {Chang},
  {Chang}, {Chen}, {Chen}, {Chen}, {Han}, {Ho}, {Huang}, {Hwang},
  {Iba{\~n}ez-Romano}, {Jiang}, {Koch}, {Kubo}, {Li}, {Lim}, {Lin}, {Liu},
  {Lo}, {Ma}, {Martin}, \& et. al.}]{Ho:2009}
{Ho} P.~T.~P. {et~al.}, 2009, ApJ, 694, 1610

\bibitem[{{H{\"o}gbom}(1974)}]{Hogbom:1974}
{H{\"o}gbom} J.~A., 1974, A\&AS, 15, 417

\bibitem[{{Holler} {et~al}\mbox{.}(2008){Holler}, {Hills}, {Jones}, {Grainge},
  \& {Kaneko}}]{Holler:2008}
{Holler} C.~M., {Hills} R.~E., {Jones} M.~E., {Grainge} K., {Kaneko} T., 2008,
  MNRAS, 384, 1207

\bibitem[{{Kawaharada} {et~al}\mbox{.}(2010){Kawaharada}, {Okabe}, {Umetsu},
  {Takizawa}, {Matsushita}, {Fukazawa}, {Hamana}, {Miyazaki}, {Nakazawa}, \&
  {Ohashi}}]{Kawaharada:2010}
{Kawaharada} M. {et~al.}, 2010, ApJ, 714, 423

\bibitem[{{LaRoque} {et~al}\mbox{.}(2006){LaRoque}, {Bonamente}, {Carlstrom},
  {Joy}, {Nagai}, {Reese}, \& {Dawson}}]{LaRoque:2006}
{LaRoque} S.~J., {Bonamente} M., {Carlstrom} J.~E., {Joy} M.~K., {Nagai} D.,
  {Reese} E.~D., {Dawson} K.~S., 2006, ApJ, 652, 917

\bibitem[{{Lemze} {et~al}\mbox{.}(2008){Lemze}, {Barkana}, {Broadhurst}, \&
  {Rephaeli}}]{Lemze:2008}
{Lemze} D., {Barkana} R., {Broadhurst} T.~J., {Rephaeli} Y., 2008, MNRAS, 386,
  1092

\bibitem[{{Lemze} {et~al}\mbox{.}(2009){Lemze}, {Broadhurst}, {Rephaeli},
  {Barkana}, \& {Umetsu}}]{Lemze:2009}
{Lemze} D., {Broadhurst} T., {Rephaeli} Y., {Barkana} R., {Umetsu} K., 2009,
  ApJ, 701, 1336

\bibitem[{{Liao} {et~al}\mbox{.}(2010){Liao}, {Proty Wu}, {Ho}, {Locutus
  Huang}, {Koch}, {Lin}, {Liu}, {Molnar}, {Nishioka}, {Umetsu}, {Wang},
  {Altamirano}, {Birkinshaw}, {Chang}, \& et. al.}]{Liao:2010}
{Liao} Y. {et~al.}, 2010, ApJ, 713, 584

\bibitem[{{Limousin} {et~al}\mbox{.}(2007){Limousin}, {Richard}, {Jullo},
  {Kneib}, {Fort}, {Soucail}, {El{\'{\i}}asd{\'o}ttir}, {Natarajan}, {Ellis},
  {Smail}, {Czoske}, {Smith}, {Hudelot}, {Bardeau}, {Ebeling}, {Egami}, \&
  {Knudsen}}]{Limousin:2007}
{Limousin} M. {et~al.}, 2007, ApJ, 668, 643

\bibitem[{{Marrone} {et~al}\mbox{.}(2009){Marrone}, {Smith}, {Richard}, {Joy},
  {Bonamente}, {Hasler}, {Hamilton-Morris}, {Kneib}, {Culverhouse},
  {Carlstrom}, {Greer}, {Hawkins}, {Hennessy}, {Lamb}, {Leitch}, {Loh},
  {Miller}, \& et. al.}]{Marrone:2009}
{Marrone} D.~P. {et~al.}, 2009, ApJL, 701, L114

\bibitem[{{Mason} {et~al}\mbox{.}(2003){Mason}, {Pearson}, {Readhead},
  {Shepherd}, {Sievers}, {Udomprasert}, {Cartwright}, {Farmer}, {Padin},
  {Myers}, {Bond}, {Contaldi}, {Pen}, {Prunet}, {Pogosyan}, {Carlstrom},
  {Kovac}, {Leitch}, \& et. al.}]{Mason:2003}
{Mason} B.~S. {et~al.}, 2003, ApJ, 591, 540

\bibitem[{{Myers} {et~al}\mbox{.}(2003){Myers}, {Contaldi}, {Bond}, {Pen},
  {Pogosyan}, {Prunet}, {Sievers}, {Mason}, {Pearson}, {Readhead}, \&
  {Shepherd}}]{Myers:2003}
{Myers} S.~T. {et~al.}, 2003, ApJ, 591, 575

\bibitem[{{Nozawa}, {Itoh} \& {Kohyama}(1998){Nozawa}, {Itoh}, \&
  {Kohyama}}]{Nozawa:1998}
{Nozawa} S., {Itoh} N., {Kohyama} Y., 1998, ApJ, 508, 17

\bibitem[{{O'Sullivan} {et~al}\mbox{.}(2008){O'Sullivan}, {Cahill}, {Murphy},
  {Gear}, {Harris}, {Ade}, {Church}, {Thompson}, {Pryke}, {Bock}, {Bowden},
  {Brown}, {Carlstrom}, {Castro}, {Culverhouse}, {Friedman}, {Ganga}, {Haynes},
  {Hinderks}, {Kovak}, {Lange}, {Leitch}, {Mallie}, {Melhuish}, {Orlando},
  {Piccirillo}, {Pisano}, {Rajguru}, {Rusholme}, {Schwarz}, {Taylor}, {Wu}, \&
  {Zemcov}}]{O'Sullivan:2007}
{O'Sullivan} C. {et~al.}, 2008, Infrared Physics and Technology, 51, 277

\bibitem[{{Padin} {et~al}\mbox{.}(2001){Padin}, {Cartwright}, {Mason},
  {Pearson}, {Readhead}, {Shepherd}, {Sievers}, {Udomprasert}, {Holzapfel},
  {Myers}, {Carlstrom}, {Leitch}, {Joy}, {Bronfman}, \& {May}}]{Padin:2001}
{Padin} S. {et~al.}, 2001, ApJL, 549, L1

\bibitem[{{Padin} {et~al}\mbox{.}(2002){Padin}, {Shepherd}, {Cartwright},
  {Keeney}, {Mason}, {Pearson}, {Readhead}, {Schaal}, {Sievers}, {Udomprasert},
  {Yamasaki}, {Holzapfel}, {Carlstrom}, {Joy}, {Myers}, \&
  {Otarola}}]{Padin:2002}
---, 2002, PASP, 114, 83

\bibitem[{{Pearson} {et~al}\mbox{.}(2003){Pearson}, {Mason}, {Readhead},
  {Shepherd}, {Sievers}, {Udomprasert}, {Cartwright}, {Farmer}, {Padin},
  {Myers}, {Bond}, {Contaldi}, {Pen}, {Prunet}, {Pogosyan}, \& et.
  al.}]{Pearson:2003}
{Pearson} T.~J. {et~al.}, 2003, ApJ, 591, 556

\bibitem[{{Peng} {et~al}\mbox{.}(2009){Peng}, {Andersson}, {Bautz}, \&
  {Garmire}}]{Peng:2009}
{Peng} E., {Andersson} K., {Bautz} M.~W., {Garmire} G.~P., 2009, ApJ, 701, 1283

\bibitem[{{QUIET Collaboration} {et~al}\mbox{.}(2010){QUIET Collaboration},
  {Bischoff}, {Brizius}, {Buder}, {Chinone}, {Cleary}, {Dumoulin}, {Kusaka},
  {Monsalve}, {N{\ae}ss}, {Newburgh}, {Reeves}, {Smith}, {Wehus}, {Zuntz},
  {Zwart}, {Bronfman}, {Bustos}, {Church}, {Dickinson}, {Eriksen}, {Ferreira},
  {Gaier}, {Gundersen}, {Hasegawa}, {Hazumi}, {Huffenberger}, {Jones},
  {Kangaslahti}, {Kapner}, {Lawrence}, {Limon}, {May}, {McMahon}, {Miller},
  {Nguyen}, {Nixon}, {Pearson}, {Piccirillo}, {Radford}, {Readhead},
  {Richards}, {Samtleben}, {Seiffert}, {Shepherd}, {Staggs}, {Tajima},
  {Thompson}, {Vanderlinde}, {Williamson}, \& {Winstein}}]{Quiet:2010}
{QUIET Collaboration} {et~al.}, 2010, preprint arXiv:1012.3191

\bibitem[{{Readhead} {et~al}\mbox{.}(2004{\natexlab{a}}){Readhead}, {Myers},
  {Pearson}, {Sievers}, {Mason}, {Contaldi}, {Bond}, {Bustos}, {Altamirano},
  {Achermann}, {Bronfman}, {Carlstrom}, {Cartwright}, {Casassus}, {Dickinson},
  {Holzapfel}, {Kovac}, {Leitch}, {May}, {Padin}, {Pogosyan}, {Pospieszalski},
  {Pryke}, {Reeves}, {Shepherd}, \& {Torres}}]{Readhead:2004a}
{Readhead} A.~C.~S. {et~al.}, 2004{\natexlab{a}}, Science, 306, 836

\bibitem[{{Readhead} {et~al}\mbox{.}(2004{\natexlab{b}}){Readhead}, {Myers},
  {Pearson}, {Sievers}, {Mason}, {Contaldi}, {Bond}, {Bustos}, {Altamirano},
  {Achermann}, {Bronfman}, {Carlstrom}, {Cartwright}, {Casassus}, {Dickinson},
  {Holzapfel}, {Kovac}, {Leitch}, {May}, {Padin}, {Pogosyan}, {Pospieszalski},
  {Pryke}, {Reeves}, {Shepherd}, \& {Torres}}]{Readhead:2004b}
---, 2004{\natexlab{b}}, Science, 306, 836

\bibitem[{{Reese} {et~al}\mbox{.}(2002){Reese}, {Carlstrom}, {Joy}, {Mohr},
  {Grego}, \& {Holzapfel}}]{Reese:2002}
{Reese} E.~D., {Carlstrom} J.~E., {Joy} M., {Mohr} J.~J., {Grego} L.,
  {Holzapfel} W.~L., 2002, ApJ, 581, 53

\bibitem[{{Schwan} {et~al}\mbox{.}(2003){Schwan}, {Bertoldi}, {Cho}, {Dobbs},
  {Guesten}, {Halverson}, {Holzapfel}, {Kreysa}, {Lanting}, {Lee}, {Lueker},
  {Mehl}, {Menten}, {Muders}, {Myers}, {Plagge}, \& et. al.}]{Schwan:2003}
{Schwan} D. {et~al.}, 2003, New Astronomy Review, 47, 933

\bibitem[{{Seljak} \& {Zaldarriaga}(1996)}]{Seljak:1996}
{Seljak} U., {Zaldarriaga} M., 1996, ApJ, 469, 437

\bibitem[{{Sievers} {et~al}\mbox{.}(2007){Sievers}, {Achermann}, {Bond},
  {Bronfman}, {Bustos}, {Contaldi}, {Dickinson}, {Ferreira}, {Jones}, {Lewis},
  {Mason}, {May}, {Myers}, {Oyarce}, {Padin}, {Pearson}, {Pospieszalski},
  {Readhead}, {Reeves}, {Taylor}, \& {Torres}}]{Sievers:2007}
{Sievers} J.~L. {et~al.}, 2007, ApJ, 660, 976

\bibitem[{{Sievers} {et~al}\mbox{.}(2003){Sievers}, {Bond}, {Cartwright},
  {Contaldi}, {Mason}, {Myers}, {Padin}, {Pearson}, {Pen}, {Pogosyan},
  {Prunet}, {Readhead}, {Shepherd}, {Udomprasert}, {Bronfman}, {Holzapfel}, \&
  {May}}]{Sievers:2003}
---, 2003, ApJ, 591, 599

\bibitem[{{Sievers} {et~al}\mbox{.}(2009){Sievers}, {Mason}, {Weintraub},
  {Achermann}, {Altamirano}, {Bond}, {Bronfman}, {Bustos}, {Contaldi},
  {Dickinson}, {Jones}, {May}, {Myers}, {Oyarce}, {Padin}, {Pearson},
  {Pospieszalski}, \& et. al.}]{Sievers:2009}
---, 2009, preprint arXiv:0901.4540v2

\bibitem[{{Skilling}(2004)}]{Skilling:2004}
{Skilling} J., 2004, in Bayesian Inference and Maximum Entropy methods in
  science and engineering: 24th International Workshop on Bayesian Inference
  and Maximum Entropy Methods in Science and Engineering, Vol. 735, pp.
  395--405

\bibitem[{{Struble} \& {Rood}(1999)}]{Struble:1999}
{Struble} M.~F., {Rood} H.~J., 1999, ApJS, 125, 35

\bibitem[{{Udomprasert} {et~al}\mbox{.}(2004){Udomprasert}, {Mason},
  {Readhead}, \& {Pearson}}]{Udomprasert:2004}
{Udomprasert} P.~S., {Mason} B.~S., {Readhead} A.~C.~S., {Pearson} T.~J., 2004,
  ApJ, 615, 63

\bibitem[{{Umetsu} \& {Broadhurst}(2008)}]{Umetsu:2008}
{Umetsu} K., {Broadhurst} T., 2008, ApJ, 684, 177

\bibitem[{{Vidal} {et~al}\mbox{.}(2011){Vidal}, {Casassus}, {Dickinson},
  {Witt}, {Castellanos}, {Davies}, {Davis}, {Cabrera}, {Cleary}, {Allison},
  {Bond}, {Bronfman}, {Bustos}, {Jones}, {Paladini}, {Pearson}, {Readhead},
  {Reeves}, {Sievers}, \& {Taylor}}]{Vidal:2011}
{Vidal} M. {et~al.}, 2011, MNRAS, 414, 2424

\bibitem[{{Watkins}(1997)}]{Watkins:1997}
{Watkins} R., 1997, MNRAS, 292, L59

\end{thebibliography}
